\documentclass[12pt, draftclsnofoot, onecolumn]{IEEEtran}
\usepackage{balance}
\usepackage{cite, graphicx,amssymb,color, pict2e, multirow, rotating}
\usepackage{graphicx,amssymb}
\usepackage[tbtags]{amsmath}
\usepackage{times,amsmath}
\usepackage{subfig}
\usepackage{flushend}
\usepackage{amsmath}
\usepackage{epsfig}
\usepackage{amssymb}
\usepackage{hyperref}
\usepackage{color}
\usepackage{psfrag}
\usepackage{subfig}
\usepackage[usenames,dvipsnames]{xcolor}
\usepackage{textcomp}
\usepackage{array}
\usepackage{tikz}
\usetikzlibrary{matrix,decorations.pathreplacing}
\usepackage{tabularx,ragged2e}
\newcolumntype{C}{>{\Centering\arraybackslash}X} 

\usepackage{algorithmic}
\usepackage{algorithm}

\newtheorem{definition}{Definition}

\newtheorem{lemma}{Lemma}

\newtheorem{theorem}{Theorem}

\title{The Study of Dynamic Caching via State Transition Field - the Case of Time-Invariant Popularity} 

\author{\IEEEauthorblockN{Jie~Gao, \IEEEmembership{Member, IEEE},   
		Lian~Zhao, \IEEEmembership{Senior Member, IEEE}, 
		and Xuemin~(Sherman)~Shen, \IEEEmembership{Fellow, IEEE} 
	}

	\thanks{
		J. Gao and X. Shen are with the Department of Electrical and Computer Engineering, University of Waterloo, Waterloo, ON, N2L 3G1, Canada (e-mail: \{jie.gao, sshen\}@uwaterloo.ca). 
		
		L. Zhao is with the Department of Electrical, Computer, and Biomedical Engineering, Ryerson University, Toronto, ON, M5B 2K3, Canada (e-mail: l5zhao@ryerson.ca).
	}
}

\begin{document}

\maketitle

\vspace{-18mm}

\begin{abstract}
	
This two-part paper investigates cache replacement schemes with the objective of developing a general model to unify the analysis of various replacement schemes and illustrate their features. To achieve this goal, we study the dynamic process of caching in the vector space and introduce the concept of state transition field (STF) to model and characterize replacement schemes. In the first part of this work, we consider the case of time-invariant content popularity based on  the independent reference model (IRM). In such case, we demonstrate that the resulting STFs are static, and each replacement scheme leads to a unique STF. The STF determines the expected trace of the dynamic change in the cache state distribution, as a result of content requests and replacements, from any initial point. Moreover, given the replacement scheme, the STF is only determined by the content popularity. Using four example schemes including random replacement (RR) and least recently used (LRU), we show that the STF can be used to analyze replacement schemes such as finding their steady states, highlighting their differences, and revealing insights regarding the impact of knowledge of content popularity. Based on the above results, STF is shown to be useful for characterizing and illustrating replacement schemes. Extensive numeric results are presented to demonstrate analytical STFs and STFs from simulations for the considered example replacement schemes.  
\end{abstract}


\vspace{-2mm}

\section{Introduction}\label{s:intro} 

Caching has been attracting an increasing amount of attention in the research of wireless communications, especially in the context of mobile edge caching \cite{J_ZPiao2019} and the joint study of communication, computation, and caching with the objective of deploying services close to mobile users \cite{J_EMarkakis2017},\cite{J_MTang2018}. The research on the performance of caching in wireless communication systems may adopt various metrics. The focus can be decreasing the content delivery latency \cite{J_SZhang2018}, alleviating congestion over the backhaul \cite{J_MEmara2018},  reducing energy consumption \cite{J_TXVu2018}, or a combination of the above~\cite{J_GLee2014}. While metrics can be different, the underlying caching performance is largely centered around one measurement, i.e., the cache hit ratio. Since a cache can only accommodate a limited portion of all contents, the cache hit ratio is determined by how the cached contents are selected and how they are updated.

Selecting the contents to be cached is relevant in the context of proactive caching. For example, an edge node can cache contents in advance during off-peak hours to reduce peak-hour network traffic load \cite{M_EBastug2014}-\cite{C_JGao2018}. The key to proactive caching is adapting to unknown content popularity or network environment, usually leading to a Markov decision problem~\cite{J_JQiao2016} or a learning problem \cite{J_SOSomuyiwa2018}.     

Updating the cached contents is relevant in the context of online caching.  Specifically, a cached content may be evicted and replaced by a new content whenever a cache miss occurs, which leads to a dynamic process that updates cached contents on the fly~\cite{J_RPedarsani2016}. The guiding rule in updating the contents is referred to as a \textit{cache replacement scheme}. Evidently, the cache replacement scheme has a significant impact on the performance of caching. In fact, even if the contents are cached proactively, cache replacement can still play an important roll in updating the cached contents while requests are being received.    

Due to the importance of cache replacement schemes, related topics have been extensively studied in various scenarios~\cite{J_SPodlipnig2003}. Classic replacement schemes include first in first out (FIFO), least recently used (LRU), least frequently used (LFU), random replacement (RR), \textit{etc.} and their variants. Some early works adopted simple probabilistic models with primitive assumptions on the request distribution \cite{J_LBelady1966} or focused on bounding the performance of the aforementioned schemes~\cite{J_DSleator1985}. More recent works adopted Markov chains to model and analyze cache replacement schemes \cite{J_GRao1978}-\cite{J_HGomaa2013}. This class of studies generally focused on deriving the steady states of the aforementioned schemes and the mixing time of their underlying Markov chains \cite{C_STarnoi2015}\cite{J_JLi2018}. In our previous work~\cite{J_JGaoTMC2018}, we considered the problem in reverse and designed the Markov chain underlying the replacement scheme so that a target set of content caching probabilities can be achieved.

Most recent works in the communications field tended to evaluate existing replacement schemes in their considered network scenarios or propose new schemes that suit their specific objectives. Chang~\textit{et~al.} studied the joint problem of cache replacement and bandwidth allocation in the scenario of peer-assisted video-on-demand systems and compared different cache replacements through simulations~\cite{J_LChang2013}. Fiore~\textit{et~al.} developed a replacement scheme for boosting content diversity in a wireless ad hoc network based on the estimated content presence at peer nodes \cite{J_MFiore2011}. A least fresh first scheme was designed in \cite{J_MMeddeb2019} to maintain the freshness of cached data for the scenario of the Internet of Things based on named data networking. Two replacement schemes were proposed for the video-on-demand service in femtocells~\cite{J_Meybodi2017}, the first of which exploits content access history for improving cache hit ratio while the second exploits information on user access delay to promote service fairness. Kamiyama~\textit{et~al.} proposed a replacement scheme for content delivery networks based on the hop count from end users to the content server with an objective to reduce network traffic load~\cite{J_NKamiyama2016}. Chattopadhyay~\textit{et~al.} investigated content replacement based on the knowledge of cached contents at neighbor base stations for a cellular network with densely deployed base stations \cite{J_AChattopadhyay2018}. A similar scenario was studied in \cite{J_ELeonardi2018}, in which the authors proposed replacement schemes that implicitly coordinate contents at caches over the network to maximize the overall hit ratio of the considered system.

While there has been abundant research on the topic of cache replacement, a model that can conveniently unify the analysis of different replacement schemes, characterize their features, and intuitively illustrate their differences is not yet available. The objective of this two-part paper is to develop such a model. Specifically, we have three targets. First, we aim to integrate cache replacement schemes under a unified general probabilistic cache replacement model and demonstrate this using several specific schemes as examples. Second, we target at studying the general cache replacement model from a novel perspective, the state transition field (STF), which characterizes replacement schemes in the vector space and captures the insights on their features. Third, we strive toward the goal of developing the model and methodology for studying cache replacements using the SFT. 

The first part of this work focuses on the case when the content popularity is time-invariant while the second part investigates the scenario of time-varying content popularity \cite{J_JGaoPartII2018}. Through the two parts of this paper, we demonstrate that a replacement scheme corresponds to a unique state transition matrix, which in turn generates a unique STF, and the resulting STF jointly determines the performance of the replacement scheme with the content request statistics. Furthermore, although such an extension is not directly included, we provide the motivation, basic model, and methodology for studying the problem in reverse:  given a performance target, can a replacement scheme be designed through determining the state transition matrix, which is in turn generated based on creating the STF according to the performance target and content request statistics?    

The contributions of the first part are the followings.    
 

First, we propose a general content replacement model based on probabilistic state transition as a unified model for cache replacement schemes. Unlike existing general model based on Markov chains, e.g., \cite{C_STarnoi2015}, we focus not just on the steady states but more on the dynamic change of the cache state distribution and describe this dynamic change in the vector space of state caching probabilities. Moreover, we introduce new ideas and results, such as the decomposition of state transition probability matrices based on contents and the mapping between state and content caching probabilities, to form a complete toolset for establishing our model. 

Second, based on the aforementioned model, we introduce STF, which is a vector field defined over the state transition domain. We demonstrate that STF can characterize and illustrate cache replacement schemes. The STF determines the expected change of the dynamic cache state distribution just like an electromagnetic field determines the movement of a charged particle placed in it (although the STF can have more than 3 dimensions). Moreover, we show that the steady state of replacement schemes can be conveniently found based on the STF. 

Third, we analyze the STF using four example replacement schemes of three types as case studies: RR, replace less popular (LP) and replace the least popular (TLP), and LRU. RR exploits no knowledge of content popularity, LP and TLP exploit perfect knowledge based on an assumption of perfect prediction, and LRU exploits imperfect knowledge from historical requests. We compare their STF and analyze the impact of the exploited knowledge on their steady states through the STF. Moreover, we conduct extensive simulations to generate the STF of the above example schemes to demonstrate the impact of replacement schemes and content popularity on the STF.    

%
%

\section{System Model}\label{s:sys} 

The scenario of $N_\mathrm{c}$ contents and a cache with size $L$ is considered. The set of all contents is denoted by $\mathcal{C}$. Without loss of generality, we assume that all contents are of identical and unit size. We do not target at a specific scenario as the model can be applied to a cache located at a small-cell base station in a cellular network, a road side unit (RSU) in a vehicular network, or even a user device for D2D caching.

\subsection{Content Request and Replacement} 

The fundamental assumption in the first part of this paper is that the requested contents at all instants, as integer random variables, are independent and identically distributed. This follows from the widely used independent reference model (IRM), a simplification of the actual request process that can be accurate with a large number of requesting users \cite{C_STarnoi2015} or within a short time frame  \cite{J_GSPaschos2018}. As the requested content follows a distribution that is time-invariant, the probability of content $l \in \mathcal{C}$ being requested can be denoted by $\upsilon_l$. The probabilities $\{\upsilon_l\}_{\forall l}$ are organized into the request probability vector $\boldsymbol{\upsilon}$  and referred to as the content popularity.  

If content $l$ is requested but not being cached, it will be downloaded and, depending on the replacement scheme, may replace one cached content. It is assumed that the download and replacement can be completed before the next content request arrives at the cache. 

\begin{figure}[tt]
	\centering {\includegraphics[width=0.48\textwidth]{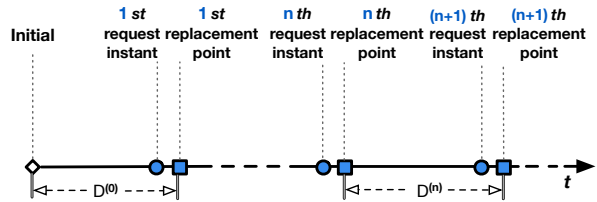}}
	\caption{Illustration of the timeline model. $\mathrm{D}^{(n)}$ represents the duration between the $n$th and the $(n+1)$th replacement points. }\label{f:sys}
\end{figure}

The timeline of the considered dynamic caching is illustrated in Fig.~\ref{f:sys}. For simplicity of notation, we place a replacement point after each request regardless of whether a replacement actually happens or not. If a replacement occurs following the $n$th request, it is completed by the $n$th replacement point.   
   
\subsection{Cache State}

\begin{figure}[tt]
	\centering {\includegraphics[width=0.40\textwidth]{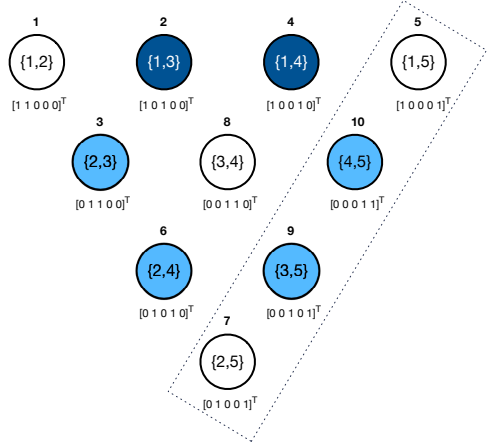}}
	\hspace{-5mm}
	\caption{An illustration of states with $N_\mathrm{c}=5$ and $L = 2$. Each circle represents a state. The number above a circle represents the state ID, and the set inside a circle represents the set of cached contents in that state. For example, state~7 caches contents~2~and~5.}\label{f:StateModel}
\end{figure}

The \emph{cache state} is introduced to describe the combination of cached contents. There are $N_\mathrm{s} = \binom{N_\mathrm{c}}{L}$ different possible combinations of cached contents, corresponding to $N_\mathrm{s}$ caching states. The set of all cache states is denoted by $\mathcal{S}$. The set of contents cached in state $k$ is denoted by $\mathcal{C}_k$. The cache state vector for state $k$ is defined as a $N_\mathrm{c} \times 1$ vector with elements determined as follows:
\begin{align}
 	\mathbf{s}_k(l) 
 	 =   \left\{
 	\begin{array}{ll}
 	1, \quad \text{if} \;\; l \in \mathcal{C}_k,  \\
 	0, \quad \text{if} \;\; l \notin \mathcal{C}_k, \\
 	\end{array}
 	\right. \forall  l\in \mathcal{C}, \forall k \in \mathcal{S},
\end{align}
where the $l$th element of vector $\mathbf{s}_k$ corresponds to the $l$th content. An example with $N_\mathrm{c}=5$ and $L = 2$ is illustrated in Fig.~\ref{f:StateModel}. In this example, there are $\binom{5}{2} = 10$ states. Each circle in the figure represents a state, while the number above the circle represents the state ID. The set given in the circle of state $k$ is the set of cached contents in state $k$, i.e., $\mathcal{C}_k$, and the vector beneath state $k$ is $\mathbf{s}_k$. For example, state~7 caches contents $\{2, 5\}$ and is represented by the cache state vector $\mathbf{s}_7 = [0\, 1\, 0\, 0\, 1]^\mathrm{T}$, where $\cdot^\mathrm{T}$ stands for transpose, given beneath the circle of state~7 in Fig.~\ref{f:StateModel}.

A state is a neighbor of state $k$ if its cached contents differ from those cached in state $k$ by just one element. The set of neighbors of state $k$ is denoted as $\mathcal{H}_k$. For any content $l \notin \mathcal{C}_k$, a content-$l$ neighbor of state $k$ is a neighboring state that caches $l$. The set of content-$l$ neighboring states of state $k$ is denoted as $\mathcal{H}_{k, l}$. Using Fig.~\ref{f:StateModel} and state 8 as an example, $\mathcal{H}_8$ is the set of all colored states, and $\mathcal{H}_{8, 1}$ is the two states with deep color.

\subsection{State and Content Caching Probabilities}

The cached contents and the cache state remain constant in the durations between consecutive replacement points (shown in Fig.~\ref{f:sys}). The state caching probability (SCP) for state $k$ and the $n$th duration, denoted by $\eta_k^{(n)}$, is the probability that the cache is in state $k$ in the $n$th duration. The content caching probability (CCP) for content $l$ and the $n$th duration, denoted by $\lambda_l^{(n)}$, is the probability that the content $l$ is cached in the $n$th duration. 

Define the SCP vector $\boldsymbol{\eta}^{(n)}$ and CCP vector $\boldsymbol{\lambda}^{(n)}$ such that $\boldsymbol{\eta}^{(n)}(k) = \eta_k^{(n)}$ and $\boldsymbol{\lambda}^{(n)}(l) = \lambda_l^{(n)}$. Evidently, $\mathbf{1}^\mathrm{T} \boldsymbol{\eta}^{(n)} = 1$ and $\mathbf{1}^\mathrm{T} \boldsymbol{\lambda}^{(n)} = L$. Based on the time line in Fig.~\ref{f:sys},  the SCP and the CCP vectors at the instant of the $(n+1)$th request are $\boldsymbol{\eta}^{(n)}$ and $\boldsymbol{\lambda}^{(n)}$, respectively.    

The SCP and the CCP are connected through cache states. Using Fig.~\ref{f:StateModel} as an example, the probability that content 5 is cached is equal to the sum of the probabilities that the states in the dotted box are cached. Define a cache state matrix  $\mathbf{C}_\mathrm{s} = [\mathbf{s}_1, \dots, \mathbf{s}_{N_\mathrm{s}}]$. In general, the relation between the SCP $\boldsymbol{\eta}^{(n)}$ and CCP $\boldsymbol{\lambda}^{(n)}$ is given by:
\begin{align}
\boldsymbol{\lambda}^{(n)} = \mathbf{C}_\mathrm{s} \boldsymbol{\eta}^{(n)}.
\end{align}


\subsection{Cache Hit Probability}

Given that content $l$ is being requested at the $(n+1)$th request, the conditional instantaneous cache hit probability is $\lambda_l^{(n)}$. The instantaneous cache hit probability at the $(n+1)$th request, denoted by $\gamma^{(n+1)}$ is given by:
\begin{align}\label{e:cacheHitDef}
\gamma^{(n+1)} = \boldsymbol{\upsilon}^\mathrm{T} \boldsymbol{\lambda}^{(n)}.
\end{align}

The symbols used in this paper are listed in Table~\ref{t:Notation}. Throughout the paper, we use lower-case bold letters for vectors, upper-case bold letters for matrices, and calligraphic letters for sets. The superscript $(\cdot)^{(n)}$ is used on letters related to the $n$th request or replacement. Greek letters are used to represent various probabilities. The indexes $m$ and $k$ are used to denote cache states, while the indexes $l$ and $q$ are used to denote contents.

\begin{table}[t!]
	\begin{center}
		\caption{List of Symbols}\label{t:Notation}
		{\setlength{\extrarowheight}{1.5pt}
			\begin{tabularx}{0.95\textwidth}{c|X}\hline\hline
				$N_\mathrm{c}$   & The number of all contents \\ \hline
				$N_\mathrm{s}$ & The number of all cache states \\ \hline
				$L$   & The cache size limit \\ \hline
				$\mathcal{C}$   & The set of all contents, i.e., $\{1, \dots, N_\mathrm{c}\}$ \\ \hline
				$\mathcal{S}$ & The set of all cache states, i.e., $\{1, \dots, N_\mathrm{s}\}$ \\ \hline
				$\mathcal{S}_l$ & The set of all cache states that cache content $l$ \\ \hline
				$\mathbf{s}_k$ & The $k$th cache state vector \\ \hline 
				$\mathcal{C}_k$  & The set of contents cached in state $k$ \\ \hline
				$\mathbf{C}_\mathrm{s}$ & The cache state matrix, i.e., $[\mathbf{s}_1, \dots, \mathbf{s}_{N_\mathrm{s}}]$ \\ \hline
				$\mathcal{H}_{k}$ & The set of all neighbors of state $k$ \\ \hline	
				$\mathcal{H}_{k, l}$ & The set of all content-$l$ neighbors of state $k$ \\ \hline
				$e(k, m)$ & The unique element in the set $\mathcal{C}_k - \mathcal{C}_m$, where $m \in \mathcal{H}_{k}$\\ \hline				
				$\upsilon_{l}$ & The request probability of content $l$ \\ \hline
				$\boldsymbol{\upsilon}$ & The content request probability vector, i.e.,  $[\upsilon_{1}, \dots, \upsilon_{N_\mathrm{c}}]^\mathrm{T}$ \\ \hline
				$\phi_{l,q, k}$ & The conditional probability that content $l$ replaces content $q$ given that cache is in state $k$ and content $l$ is requested \\ \hline
				$\boldsymbol{\Theta}$ & The state transition probability matrix \\ \hline
				$\boldsymbol{\Theta}_l$ & The conditional state transition probability matrix given that content $l$ is requested\\ \hline
				$\boldsymbol{\Theta}(m, k)$ & The probability of transitioning from state $k$ to state $m$ \\ \hline	
				$\boldsymbol{\Theta}_l(m, k)$ & The probability of transitioning from state $k$ to state $m$ given that content $l$ is requested\\ \hline			
				$\eta_{k}^{(n)}$ & The SCP for state $k$ in the duration from the $n$th to the $(n+1)$th replacement \\ \hline
				$\boldsymbol{\eta}^{(n)}$ &  The SCP vector in the duration from the $n$th to the $(n+1)$th replacement,  i.e, $[\eta_{1}^{(n)}, \dots, \eta_{N_\mathrm{s}}^{(n)}]^\mathrm{T}$ \\ \hline
				$\lambda_l^{(n)}$ & The CCP for content $l$ in the duration from the $n$th to the $(n+1)$th replacement  \\ \hline
				$\boldsymbol{\lambda}^{(n)}$ & The CCP vector in the duration from the $n$th to the $(n+1)$th replacement, i.e., $[\lambda_{1}^{(n)}, \dots, \lambda_{N_\mathrm{c}}^{(n)}]^\mathrm{T}$ \\ \hline
				$\gamma^{(n)}$ & The instantaneous cache hit probability at the $n$th request \\ \hline
				$\mathbf{u}(\boldsymbol{\eta})$ & The state transition field at $\boldsymbol{\eta}$ \\ \hline
				$\mathbf{u}_l(\boldsymbol{\eta})$ & The content-$l$ state transition field at $\boldsymbol{\eta}$ \\ \hline
				$u_{m, l}(\boldsymbol{\eta})$ & The $m$th element of the state transition field at $\boldsymbol{\eta}$ \\ \hline
				$u_{m, l}(\boldsymbol{\eta})$ & The $m$th element of the content-$l$ state transition field at $\boldsymbol{\eta}$ \\ 
				\hline\hline
			\end{tabularx}
		}
	\end{center}
	\vspace{-0.2cm}
\end{table}

\section{General Content Replacement Model and State Transition Field}

If the cache is at state $k$ while content $l \notin \mathcal{C}_k$ is requested, the cache downloads content $l$ and decides whether to replace a cached content with content $l$. In the general model, the probability of replacing content $q$ with content $l$ when the cache is at state $k$ is denoted by $\phi_{l, q, k}$, for any $q \in\mathcal{C}_k$ and $l \notin \mathcal{C}_k$. For each state, there are $ L (N_\mathrm{c}- L)$ possible replacements.

\subsection{General Cache State Transition Model}

A content replacement triggers a cache state transition. For neighboring states $k$ and $m$ which satisfies $m \in \mathcal{H}_{k,l}$ and $k \in \mathcal{H}_{m, q}$, replacing content $q$ with $l$ triggers a transition from state $k$ to state $m$. The conditional cache state transition probabilities given that content $l$ is requested can be organized into the following matrix $\boldsymbol{\Theta}_l$:
\begin{align}
\boldsymbol{\Theta}_l(m, k) 
\! = \!  \left\{
\begin{array}{ll}
1, & \text{if} \,\;   k = m \; \text{and} \; l \in \mathcal{C}_k, \\
1\! - \!\!\! \sum\limits_{m^\prime \in \mathcal{H}_{k,l}} \phi_{l, e(k, m^\prime), k}, & \text{if} \,\;  k = m \; \text{and} \;  l \notin \mathcal{C}_k, \\
\phi_{l, e(k, m), k}, &\text{if} \,\; m \in \mathcal{H}_{k,l}, \\ 
0, & \text{otherwise}, \\
\end{array}
\right.
\end{align}
where $e(k, m)$ denotes the unique content that is cached by state $k$ but not state $m$ given that $k \in \mathcal{H}_{m}$. Accordingly, the overall cache state transition probability matrix in the general case is given by:
\begin{align}\label{e:OverallSTP}
\boldsymbol{\Theta} =  \sum\limits_{l \in \mathcal{C}} \upsilon_l \boldsymbol{\Theta}_l.
\end{align} 

From the definition of the SCP vector $\boldsymbol{\eta}^{(n)}$ and state transition probability matrix $\boldsymbol{\Theta} $, it can be seen that:
\begin{align}\label{e:etaChange1Step}
\boldsymbol{\eta}^{(n)} = \boldsymbol{\Theta} \boldsymbol{\eta}^{(n-1)}. 
\end{align} 

It is worth mentioning that the model can be extended to the scenario in which each content request (and replacement) involves multiple contents. In such case, assuming that each request is for a block of $B$ contents ($B<L$), there are $N_\mathrm{B} = \binom{N_\mathrm{c}}{B}$ different blocks. Then, eq.~\eqref{e:OverallSTP} can be extended as follows:
\begin{align}
\boldsymbol{\Theta}^{\mathrm{B}} =  \sum\limits_{b = 1}^{N_\mathrm{B}} \upsilon_b^{\mathrm{B}} \boldsymbol{\Theta}_b^{\mathrm{B}},
\end{align}
where $\upsilon_b^{\mathrm{B}}$ is the probability that the $b$th block is requested, and $\boldsymbol{\Theta}_b^{\mathrm{B}}$ is the conditional cache state transition probabilities given that block $b$ is requested. The size of $\boldsymbol{\Theta}_b^{\mathrm{B}}$ remains $N_\mathrm{s} \times N_\mathrm{s}$. However, for any given state, e.g., state $k$, the set of its neighbors $\mathcal{H}_k$ will contain more states under block replacement, and the set of its content-$l$ neighbors $\mathcal{H}_{k, l}$ will be replaced by a set of block-$b$ neighbors. The extension is straightforward and the details are omitted here.     

\subsection{STF}

Denote the general SCP without specifying any time instant as $\boldsymbol{\eta}$. Consider $\boldsymbol{\eta}$ as a point in the $N_\mathrm{s}$-dimensional vector space. Driving by the requests and replacements, $\boldsymbol{\eta}$ varies in the following domain: 
\begin{align}
\mathcal{D} \!= \left\{\!\big(\eta_1, \dots, \eta_{N_\mathrm{s}}\big) \bigg| 0 \leq \eta_k \leq 1, \forall k\in \mathcal{S};\, \sum\limits_k \eta_k = 1 \right\}.
\end{align}  

The expected `movement' of $\boldsymbol{\eta}$ in $\mathcal{D}$ after the $n$th replacement point, assuming a replacement actually happens, is characterized by $\boldsymbol{\eta}^{(n)} - \boldsymbol{\eta}^{(n-1)}$. This difference, in turn, is determined by three factors:
\begin{itemize}
	\item the current position of $\boldsymbol{\eta}$ in $\mathcal{D}$, i.e., the value of $\boldsymbol{\eta}^{(n-1)}$
	\item the content popularity $\boldsymbol{\upsilon}$
	\item the state transition probability matrix $\boldsymbol{\Theta}$,
\end{itemize}
while $\boldsymbol{\Theta}$ is determined by the replacement scheme and generally dependent on $\boldsymbol{\upsilon}$ (and such dependence is shown in eq.~\eqref{e:OverallSTP}).   

Define the STF at the point $\boldsymbol{\eta}^{(n-1)}$ using the aforementioned difference:
\begin{align}\label{e:StateTranFielddef}
\mathbf{u}(\boldsymbol{\eta}^{(n-1)}) = \boldsymbol{\eta}^{(n)} - \boldsymbol{\eta}^{(n-1)}.
\end{align}
Substituting eq.~\eqref{e:etaChange1Step} into eq.~\eqref{e:StateTranFielddef}, it follows that:
\begin{align}
\mathbf{u}(\boldsymbol{\eta}^{(n-1)})  =  \boldsymbol{\Theta} \boldsymbol{\eta}^{(n-1)} - \boldsymbol{\eta}^{(n-1)}.
\end{align}

The STF is a vector field defined over the domain $\mathcal{D}$. It can be seen that understanding the STF can provide insight into the design and performance analysis of replacement schemes. Similar to a magnetic or electric field, the STF can vary in direction and strength at different points in the domain (although the STF exists mathematically but not physically). 

\begin{figure}[tt]
	\centering {\includegraphics[width=0.37\textwidth]{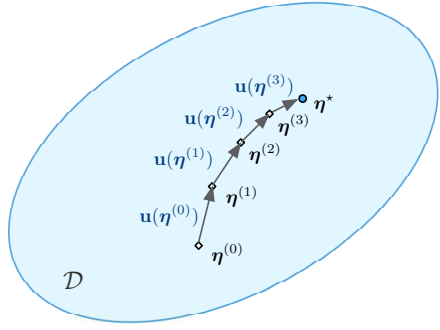}}
	\vspace{-0mm}
	\caption{An illustration of STF at four points, i.e., $\boldsymbol{\eta}^{(0)}$ to $\boldsymbol{\eta}^{(3)}$. The end point $\boldsymbol{\eta}^{\star}$ represents the steady state, at which the STF diminishes to an all-zero vector. }\label{f:StateTranField}
\end{figure}

In the definition eq.~\eqref{e:StateTranFielddef}, the $\boldsymbol{\eta}^{(n-1)}$ in the brackets specifies a point in the domain $\mathcal{D}$. If the STF is known at all points in $\mathcal{D}$, then a path can be identified from any initial point, as illustrated in Fig.~\ref{f:StateTranField}, the end of which gives the steady state of the replacement scheme while the number of steps in the path reflects the time for the underlying Markov chain to attain its stationary state from that initial point. Different replacement schemes yield different STFs, and the impact is conveyed through $\boldsymbol{\Theta}$. Therefore, the STF is a complete characterization of replacement schemes.     

It is worth noting that the STF does not change over time under the IRM in general, as $\boldsymbol{\upsilon}$ and $\boldsymbol{\Theta}$ are both constant.

\subsection{Content-specific STF}

The STF can be decomposed. Define:
\begin{align}\label{e:u_l}
\mathbf{u}_l(\boldsymbol{\eta}^{(n-1)}) =  \boldsymbol{\Theta}_l \boldsymbol{\eta}^{(n-1)} - \boldsymbol{\eta}^{(n-1)}. 
\end{align}
It follows that:
\begin{align}\label{e:u_Decomp}
\sum\limits_{l \in \mathcal{C}} \! \upsilon_l \mathbf{u}_l(\boldsymbol{\eta}^{(n-1)}) \!=\!   \sum\limits_{l \in \mathcal{C}} \upsilon_l \boldsymbol{\Theta}_l \boldsymbol{\eta}^{(n-1)} \!-\! \boldsymbol{\eta}^{(n-1)}  \!=\!  \mathbf{u}(\boldsymbol{\eta}^{(n-1)}), 
\end{align}
where the last step uses eq.~\eqref{e:OverallSTP}. Accordingly, $\mathbf{u}_l(\boldsymbol{\eta}^{(n-1)})$ can be considered as the content-specific STF that represents the `movement' of $\boldsymbol{\eta}$ from the point $\boldsymbol{\eta}^{(n-1)}$ after content $l$ is requested. The superposition of all content-specific STFs, weighted by the corresponding content popularity, yields the overall STF. 

It is not difficult to see that the following equalities hold:
\begin{align}
\mathbf{1}^\mathrm{T}\mathbf{u}_l(\boldsymbol{\eta}^{(n-1)}) &= 0, \; \forall l \in \mathcal{C}, \, \forall \boldsymbol{\eta}^{(n-1)} \in \mathcal{D} \\
\mathbf{1}^\mathrm{T}\mathbf{u}(\boldsymbol{\eta}^{(n-1)}) &= 0, \; \forall \boldsymbol{\eta}^{(n-1)} \in \mathcal{D}.
\end{align}

\section{State Transition Matrices of Specific Replacement Schemes}\label{s:STMatrices}

In this section, we demonstrate how four specific replacement schemes, i.e., RR, LP, TLP, and LRU, fit into the general content replacement model in the preceding section. As the impact of replacement schemes on the STFs is conveyed through the state transition matrix $\boldsymbol{\Theta}$, the focus will be on finding $\boldsymbol{\Theta}$ for the considered schemes. 

The four replacement schemes can be categorized into three groups based on the content popularity information that they exploit. 
\begin{itemize}
\item RR does not use any content popularity information;
\item Both LP and TLP rely on the prediction of content popularity, and a perfect prediction will be assumed.
\item LRU exploits imperfect content popularity information from request history, i.e., the information of recent content requests.
\end{itemize}
The impact of the difference in the exploited content popularity information on the STF will be presented in subsequent sections of this paper.   

\subsection{RR}

For RR, the conditional content replacement probability $\phi_{l, q, k}$ reduces to a constant:
\begin{align}
\phi_{l, q, k} = \phi \in (0, 1/L], \quad \forall q\in \mathcal{C}_k, l \notin \mathcal{C}_k.
\end{align}

Accordingly, the conditional state transition probability matrix $\boldsymbol{\Theta}_l$ is given by:
\begin{align}\label{e: ThetaDef}
\boldsymbol{\Theta}_{\mathrm{RR}, l}(m, k) 
=   \left\{
\begin{array}{ll}
1, & \text{if} \;\;  l \in \mathcal{C}_k  \;\;\text{and} \;\;  k = m, \\
1 - L \phi, & \text{if} \;\;  l \notin \mathcal{C}_k  \;\;\text{and} \;\;  k = m, \\
\phi, & \text{if} \;\; m \in \mathcal{H}_{k,l}, \\ 
0, & \text{otherwise}, \\
\end{array}
\right.
\end{align}
i.e., the probabilities of content $l$ replacing a cached content and no replacement are $L\phi$ and $1 - L\phi$, respectively.

The overall state transition probability matrix $\boldsymbol{\Theta}_\mathrm{RR}$ is given by:
\begin{align}\label{e: ThetaOverallDef}
\boldsymbol{\Theta}_\mathrm{RR} (m, k)
=   \left\{
\begin{array}{ll}
1 - L \phi \sum\limits_{l \notin \mathcal{C}_k} \upsilon_l , & \text{if} \;\;  k = m, \\
\phi \upsilon_{e(m, k)}, & \text{if} \;\; m \in \mathcal{H}_k, \\ 
0, & \text{otherwise}. \\
\end{array}
\right. 
\end{align}

\subsection{LP}

Denote the predicted content popularity by $\tilde{\boldsymbol{\upsilon}}$. Using LP, the requested content $l\notin \mathcal{C}_k$ may replace a cached content $q$ in state $k$ if $\tilde{\upsilon}_l > \tilde{\upsilon}_q$, i.e., the requested content is more popular. The conditional state transition  probability is given by:   
\begin{align}\label{e:ThetalReplaceLess}
& \boldsymbol{\Theta}_{\mathrm{LP},l}(m, k) \nonumber \\
& =\! \left\{
\begin{array}{ll}
\! 1, & \text{if} \;  l \in \mathcal{C}_k  \;\;\text{and} \;\;  k = m, \\
\! 1 - \alpha, & \text{if} \;  l \notin \mathcal{C}_k, \;  k = m, \;\text{and} \; \tilde{\upsilon}_l > \tilde{\upsilon}_q, \\
\! \alpha \phi_{l,q, k}, & \text{if} \; m \in \mathcal{H}_{k,l}, k \in \mathcal{H}_{m,q}, \, \text{and}\; \tilde{\upsilon}_l > \tilde{\upsilon}_q,\\ 
\! 0, & \text{otherwise}, \\
\end{array}
\right.  
\end{align}
where $\alpha$ is a parameter that controls the probability of a replacement. 

The conditional replacement probability, assuming that $\tilde{\upsilon}_l > \tilde{\upsilon}_q$, is set to be proportional to $\tilde{\upsilon}_l - \tilde{\upsilon}_q$, as follows: 
\begin{align}
\phi_{l,q, k} = \frac{\tilde{\upsilon}_l - \tilde{\upsilon}_q}{ \sum\limits_{t \in \mathcal{C}_{k, l}^\downarrow} (\tilde{\upsilon}_l - \tilde{\upsilon}_t) },
\end{align}
where
\begin{align}
\mathcal{C}_{k, l}^\downarrow = \{t| t \in \mathcal{C}_k, \tilde{\upsilon}_{t} < \tilde{\upsilon}_{l} \}.
\end{align}

Order the states based on $\sum_{t\in \mathcal{C}_k} \tilde{\upsilon}_t$, i.e., the summation of the predicted content request probability of each state, in a non-decreasing order. Then, it can be shown that the state transition matrix $\boldsymbol{\Theta}_\mathrm{LP}$ becomes a lower-triangular matrix:
\begin{align}\label{e:ThetaOverallDefLES}
& \boldsymbol{\Theta}_\mathrm{LP}(m, k) \nonumber \\
& = \left\{
\begin{array}{ll}
\sum\limits_{q \in \mathcal{C}_{k}} \!\upsilon_{q} \!+\!\! \sum\limits_{l \in \bar{\mathcal{C}}^\downarrow_{k}} \!\upsilon_{l} \!+\!\! \sum\limits_{l \in \bar{\mathcal{C}}^\uparrow_{k}} \!\upsilon_{l} (1 \!-\! \alpha), \!\!& \text{if} \;  m = k, \\
\alpha \upsilon_{e(m, k)} \phi_{e(m,k), e(k,m), k} , & \text{if} \; m \!>\! k \; \text{and} \; m \!\in\! \mathcal{H}_k, \\
0, &\text{otherwise}. \\
\end{array}
\right.  
\end{align}  
in which
\begin{subequations}
\begin{align}
\bar{\mathcal{C}}^\downarrow_{k} = \left\{l \mid l \notin \mathcal{C}_{k}, \tilde{\upsilon}_l \leq \min_{t \in \mathcal{C}_k} \{ \tilde{\upsilon}_t\} \right\},  \\
\bar{\mathcal{C}}^\uparrow_{k} = \left\{l \mid l \notin \mathcal{C}_{k}, \tilde{\upsilon}_l > \min_{t \in  \mathcal{C}_k} \{ \tilde{\upsilon}_t\} \right\}.
\end{align}
\end{subequations}

\subsection{TLP}

Denote the least popular content of state $k$ based on the prediction by $q^\dagger(k)$, i.e., 
\begin{align}\label{e:qdaggerLEA}
q^\dagger(k) = \mathop{\mathrm{argmin}}\limits_{t \in \mathcal{C}_k} \{\tilde{\upsilon}_t\}. 
\end{align}
Using TLP, the requested content $l\notin \mathcal{C}_k$ can only replace $q^\dagger(k)$ when the cache is in state $k$, and the replacement can happen only if $\tilde{\upsilon}_l > \tilde{\upsilon}_{q^\dagger(k)}$. The conditional state transition  probability is given by:   
\begin{align}\label{e:ThetalReplaceLeast}
& \boldsymbol{\Theta}_{\mathrm{TLP},l}(m, k) \nonumber \\
& =\! \left\{
\begin{array}{ll}
\!\!1, &\!\!\! \text{if} \;  l \in \mathcal{C}_k  \,\text{and} \;  k = m, \\
\!\!1 \!- \phi_{l,q^\dagger(k), k}, &\!\!\! \text{if} \;  l \notin \mathcal{C}_k  \, \text{and} \;  k = m, \,\text{and} \; \tilde{\upsilon}_l \!>\! \tilde{\upsilon}_{q^\dagger(k)},\\
\!\!\phi_{l,q^\dagger(k), k}, &\!\!\! \text{if} \; m \!\in\! \mathcal{H}_{k, l}, k \!\in\! \mathcal{H}_{m, q^\dagger(k)}, \text{and} \; \tilde{\upsilon}_l \!>\! \tilde{\upsilon}_{q^\dagger(k)}, \\ 
\!\!0, &\!\!\! \text{otherwise}. \\
\end{array}
\right.  
\end{align}

Two choices of the replacement probability $\phi_{l, q^\dagger(k), k}$ are considered when $\tilde{\upsilon}_l > \tilde{\upsilon}_{q^\dagger(k)}$: $\phi_{l, q^\dagger(k), k} = 1$  and $\phi_{l, q^\dagger(k), k} = \tilde{\upsilon}_{l} - \tilde{\upsilon}_{q^\dagger(k)}$. In the first case, the replacement always occurs, and the TLP in such case will be referred to as TLP-A. In the second case, the replace occurs probabilistically, and the the TLP in such case will be referred to as TLP-P. Intuitively, TLP-A would lead to faster convergence while TLP-P could be useful when each replacement incurs a replacement cost.   

Order the states based on $\sum_{t\in \mathcal{C}_k} \tilde{\upsilon}_t$, i.e., the summation of the predicted content request probability of each state, in a non-decreasing order. Then,  the state transition matrix $\boldsymbol{\Theta}_\mathrm{TLP}$ also becomes a lower-triangular matrix:
\begin{align}\label{e: ThetaOverallDefLEA}
& \boldsymbol{\Theta}_\mathrm{TLP}(m, k) \nonumber \\
& =\! \left\{\!\!
\begin{array}{ll}
\sum\limits_{q \in \mathcal{C}_{k}} \upsilon_{q} \!+\! \sum\limits_{l \in \bar{\mathcal{C}}^\downarrow_{k}} \upsilon_{l} + \sum\limits_{ l \in \bar{\mathcal{C}}^\uparrow_{k}} \upsilon_{l} (1-  \phi_{l,q^\dagger(k), k}),  &\text{if} \,  m = k, \\
\upsilon_{e(m, k)} \phi_{e(m,k),q^\dagger(k), k}, \;  &\hspace{-2.8cm}\text{if} \; m \!>\! k \;\text{and}\; e(k, m) = q^\dagger(k), \\
0, \;  &\text{otherwise}. \\
\end{array}
\right.  
\end{align}

\subsection{LRU}


When the cache is in state $k$ while content $l$ is requested,  the conditional state transition probability matrix $\boldsymbol{\Theta}_l$ is given by:
\begin{align}\label{e: ThetaDefLRU}
\boldsymbol{\Theta}_{\mathrm{LRU}, l}(m, k) 
=   \left\{
\begin{array}{ll}
1, \; & \text{if} \;\;  l \in \mathcal{C}_k  \;\;\text{and} \;\;  k = m,  \\
\rho^\text{LRU}_{e(k,m)|k},\; & \text{if} \;\; m \in \mathcal{H}_{k, l}, \\ 
0, \; & \text{otherwise}. \\
\end{array}
\right.
\end{align}
where $\rho^\text{LRU}_{e(k,m)|k}$ represents the conditional probability that content $e(k, m)$ is the least recently used content given that the cache is in state $k$. 
The probability $\rho^\text{LRU}_{e(k,m)|k}$ can be found, as a simplified special case under IRM, based on Lemma~1 in the second part of this two-part paper, which addresses the more general case of time-varying content popularity \cite{J_JGaoPartII2018}. 

The overall state transition probability matrix $\boldsymbol{\Theta}_\mathrm{LRU}$ is given by:
\begin{align}\label{e: ThetaOverallDefLRU}
\boldsymbol{\Theta}_\mathrm{LRU} (m, k)
=   \left\{
\begin{array}{ll}
\sum\limits_{l \in \mathcal{C}_k} \upsilon_l, \; & \text{if} \;\;  k = m, \\
\upsilon_{e(m, k)} \rho^\mathrm{LRU}_{e(k, m)|k}, \;& \text{if} \;\; m \in \mathcal{H}_k, \\ 
0, \; & \text{otherwise}. \\
\end{array}
\right. 
\end{align}

Note that, unlike RR and LRU, LP and TLP are not practical replacement schemes. However, the latter two are considered here for the purpose of analyzing what the STF of a replacement scheme would become in the ideal case with perfect content popularity information, as a comparison to the cases with no and imperfect content popularity information (e.g., RR and LRU, respectively).

\section{STF based Analysis for Cache Replacement Under Time-invariant Content Popularity}\label{s:Analysis} 

In this section, we analyze specific replacement schemes using the STF to demonstrate that analysis based on STF can characterize the features of different replacement schemes and reveal insights regarding their steady states.

\subsection{RR}

Using the definition of content-specific STF in eq.~\eqref{e:u_l} and the state transition probability matrix of RR in eq.~\eqref{e: ThetaDef}, it can be shown that the $m$th element of the content-specific STF at $\boldsymbol{\eta}$ is given by:
\begin{align}\label{e:ul_RR}
u_{m, l, \mathrm{RR}} (\boldsymbol{\eta}) =  \left\{
\begin{array}{ll}
\phi \sum\limits_{\{k| m \in \mathcal{H}_{k,l}\}} \eta_k, \; &\text{if}\;  l \in \mathcal{C}_m, \\ 
-L \phi \eta_m, \;\; &\text{otherwise}. 
\end{array}
\right.
\end{align}

Using the STF in eq.~\eqref{e:ul_RR}, the following result becomes straightforward.
\begin{theorem}\label{t:PropertyRRsteady}
	The steady state of RR, denoted by $ \boldsymbol{\eta}^\star$,  is independent on the parameter $\phi$ and satisfies the following property:
		\begin{align}
		\eta_m^\star \sum\limits_{l\notin \mathcal{C}_m} \upsilon_l =   \frac{1}{L}\sum\limits_{k \in \mathcal{H}_m }  \eta_{k}^\star \upsilon_{e(m, k)}, \forall m \in \mathcal{S}.  \label{e:etaSteadyRNProp2}
		\end{align}
\end{theorem}

\textit{Proof}: See Section~\ref{p:PropertyRRsteady} in Appendix. 

The property in Theorem~\ref{t:PropertyRRsteady} can be used to obtain a closed-form expression of the steady state. Define $N_\mathrm{s}$ vectors, one for each state, so that   
\begin{align}
\mathbf{a}_m (k) = \left\{
\begin{array}{ll}
\sum\limits_{l\notin \mathcal{C}_m} \upsilon_l, \; &\text{if}\;  k = m, \\ 
- \frac{1}{L} \upsilon_{e(m, k)}, \; &\text{if}\;  m \in \mathcal{H}_{k}, \\ 
0, \;\; &\text{otherwise}. 
\end{array}
\right.
\end{align}
where $\mathbf{a}_m (k)$ represents the $k$th element of the vector for the $m$th state. Then, $N_\mathrm{s}-1$ out of the $N_\mathrm{s}$ vectors are linearly independent. Define matrix $\mathbf{A}$ as follows:
\begin{align}
\mathbf{A} = [\mathbf{a}_1, \dots,  \mathbf{a}_{N_\mathrm{s}-1}, \mathbf{1} ]^\mathrm{T},
\end{align}
where $\mathbf{1}$ is an all-one vector. Then, the steady state $\boldsymbol{\eta}^\star$  can be given by 
\begin{align}
\boldsymbol{\eta}^\star = \mathbf{A}^{-1} \mathbf{g},
\end{align}
in which $\mathbf{g} = [0, \dots, 0, 1]^\mathrm{T}$ is the vector that has 0 as its first $N_\mathrm{s}-1$ elements and 1 as its last element.

Evidently, the steady state of RR does not maximize cache hit probability as RR does not exploit any content popularity information. The property in Theorem~\ref{t:PropertyRRsteady} characterizes the steady state of RR. Specifically, eq.\eqref{e:etaSteadyRNProp2} shows that the steady state of RR achieves such balance that, if a randomly selected cached content is to be replaced by a random content not cached, the resulting expected cache miss probability due to this replacement should be equal to the cache miss ratio of the steady state without any replacement.

The rate of convergence of a finite-state ergodic Markov chain is decided by the second largest eigenvalue of its transition probability matrix \cite{B_DLevin2008}. Specifically, it holds that~\cite{LN_SArora}:
\begin{align}\label{e:ErgodicConvergenceResult}
\| \boldsymbol{\Theta}^t \boldsymbol{\eta}^{(0)} - \boldsymbol{\eta}^{\star}(\boldsymbol{\Theta}) \|_2 \leq d_2^t(\boldsymbol{\Theta}) \|\boldsymbol{\eta}^{(0)}\|_2 
\end{align}
for any initial state distribution $\boldsymbol{\eta}^{(0)}$, where $\boldsymbol{\Theta}$ represents any ergodic Markov chain, $\boldsymbol{\eta}^{\star}(\boldsymbol{\Theta})$ represents the corresponding steady state,  $d_2(\boldsymbol{\Theta})$ represents the second largest eigenvalue of $\boldsymbol{\Theta}$, and $t$ is the number of steps since the initial point. While it is generally impossible to derive an eigenvalue of an arbitrary transition matrix $\boldsymbol{\Theta}$ in closed-form, the bounds on the second largest eigenvalue of a reversible transition matrix can be estimated \cite{J_SGWalker2011}. The STF provides another intuitive perspective for analyzing the rate of convergence. In the case of RR, a larger $\phi$ implies stronger STF while the direction of the STF at all points remains the same. Therefore, a larger $\phi$ generally leads to a shorter mixing time.

\subsection{LP and TLP}

Note that, in practice, the $L$ most popular contents can be placed in the cache from the beginning without using LP or TLP for replacements if the content popularity is known. However, as we intend to analyze the impact of content popularity information adopted by a replacement scheme on the path of state cache distribution starting from an arbitrary state, the analysis of LP and TLP is of interest.   

For LP and TLP, the steady state is straightforward. Sort the contents based on a nondecreasing order of their predicted popularity so that $\tilde{\upsilon}_l \geq \tilde{\upsilon}_q$ if $l \geq q$. Sort the states based on $\sum_{t\in \mathcal{C}_k} \tilde{\upsilon}_t$, i.e., the summation of the predicted content request probability of each state, in a non-decreasing order. Then, the $L$ least popular contents are cached in state 1, and the $L$ most popular contents are cached in state $N_\mathrm{s}$.

\begin{lemma}\label{l:SteadyStateLESLEA}
	The steady state for both LP and TLP is $\boldsymbol{\eta}^\star = [0, \dots, 0, 1]$.
\end{lemma}

The proof is straightforward given that, for any $k \in \mathcal{S}$, the following two facts hold: 1).~State $k$ can only transition to state $m$ if $m > k$; and 2).~State $k$ transitions to at least one neighboring state in $\mathcal{H}_k$ with a positive probability. 
The two observations can be made based on eq.~\eqref{e:ThetaOverallDefLES} and eq.~\eqref{e: ThetaOverallDefLEA}. 

Compared to the steady state of RR, the result in Lemma~\ref{l:SteadyStateLESLEA} reflects the impact of exploiting content popularity information on the steady state of a replacement scheme.

Since both $\boldsymbol{\Theta}_\mathrm{LP}$ and $\boldsymbol{\Theta}_\mathrm{TLP}$ are lower-triangular matrices, the eigenvalues of  $\boldsymbol{\Theta}_\mathrm{LP}$ and $\boldsymbol{\Theta}_\mathrm{TLP}$ are their respective diagonal elements. Evidently, neither of $\boldsymbol{\Theta}_\mathrm{LP}$ and $\boldsymbol{\Theta}_\mathrm{TLP}$ is ergodic. Nevertheless, the second largest eigenvalue of both falls in $(0, 1)$ in both cases, and the result in eq.~\eqref{e:ErgodicConvergenceResult} still holds for  $\boldsymbol{\Theta}_\mathrm{LP}$ and $\boldsymbol{\Theta}_\mathrm{TLP}$. The largest eigenvalue is 1 in both cases. The second largest eigenvalue, which determines the mixing time of LP and TLP, is given by the following result. 

\begin{lemma}\label{l:2ndEigenValueLESLEA}
	The second largest eigenvalues of $\boldsymbol{\Theta}_\mathrm{LP}$ and $\boldsymbol{\Theta}_\mathrm{TLP}$ are given by
	\begin{align}
	g_2(\boldsymbol{\Theta}_\mathrm{LP}) &= 1 - \alpha \tilde{\upsilon}_{\hat{l}} \\
	g_2(\boldsymbol{\Theta}_\mathrm{TLP}) &= 1 - \tilde{\upsilon}_{\hat{l}} \phi_{\hat{l}, \hat{l} -1, N_\mathrm{s} -1} 
	\end{align}   
\end{lemma}
where $\hat{l} = N_\mathrm{c} -L + 1$. 

\textit{Proof}:  See Section~\ref{p:2ndEigenValueLESLEA} in Appendix. 

Based on Lemma~\ref{l:2ndEigenValueLESLEA}, the rate of convergence depends on $\alpha$ in the case of LP and $\phi_{\hat{l}, \hat{l} -1, N_\mathrm{s}- 1}$ in the case of TLP. Furthermore, it can also be seen from Lemma~\ref{l:2ndEigenValueLESLEA} that the rate of convergence in both cases also depends on the popularity of a particular content,  i,e., the $(N_\mathrm{c} -L + 1)$th content, or equivalently, the $L$th most popular content.

Unlike RR and LRU, which do not rely on the content popularity information, prediction error in the content request probabilities could have an impact on either the STF or both the STF and the steady state of LP and TLP. Specifically, if there are errors in the prediction but the set of the $L$ most popular contents is predicted correctly, then the predicted STF can differ from the actual STF but the steady state will not be affected. By contrast, if the predicted $L$ most popular contents are different from the actual $L$ most popular contents, then both the STF and the steady state from the prediction will differ from their respective actual values.

\subsection{LRU}

Using the definition of content-specific STF in eq.~\eqref{e:u_l} and the state transition probability matrix of LRU in eq.~\eqref{e: ThetaDefLRU}, it can be shown that the $m$th element of the content-specific STF at $\boldsymbol{\eta}$ is given by:
\begin{align}\label{e:ul_LRU}
u_{m, l, \mathrm{LRU}} (\boldsymbol{\eta}) =  \left\{
\begin{array}{ll}
\sum\limits_{ \{ k | m \in \mathcal{H}_{k,l} \} } \rho^\mathrm{LRU}_{e(k,m)|k}  \eta_k, \; & \text{if}\; l \in \mathcal{C}_m, \\
- \eta_m, \; &\text{otherwise}.\\ 
\end{array}
\right.
\end{align}

Under the IRM model, the probabilities $\{\rho^\mathrm{LRU}_{e(k,m)|k}\}_{\forall k, \forall m \in \mathcal{H}_k}$ are constants and can be calculated. Given $\{\rho^\mathrm{LRU}_{e(k,m)|k}\}$, the following result  regarding the steady state in the case of LRU can be found using the STF. 

\begin{theorem}\label{t:etaSteadyLRU}
	The steady state $\boldsymbol{\eta}^\star$ in the case of LRU satisfies the following property:
	\begin{align}\label{e:etaSteadyLRU}
	 \eta_m^\star \sum\limits_{l \notin \mathcal{C}_m}  \upsilon_l   = \sum\limits_{k \in \mathcal{H}_m }  \upsilon_{e(m, k)}  \rho^\mathrm{LRU}_{e(k,m)|k} \eta_{k}^\star.
	\end{align}
\end{theorem}

\textit{Proof}: See Section~\ref{p:etaSteadyLRU} in Appendix. 

Comparing eq.~\eqref{e:etaSteadyLRU} and eq.~\eqref{e:etaSteadyRNProp2} reveals an interesting insight. Denote the steady state SCP in the case of RR by $\boldsymbol{\eta}^{\star}_\mathrm{RR}$. The STF at the point $\boldsymbol{\eta}^{\star}_\mathrm{RR}$ in the case of LRU is given by: 
\begin{align}\label{e:compSteadyRR_LRU}
&\hspace{-2mm} u_{m, \mathrm{LRU}} (\boldsymbol{\eta}^\star_\mathrm{RR}) \nonumber \\
=&  \!\sum\limits_{l \in \mathcal{C}_m} \! \upsilon_l \cdot u_{m, l, \mathrm{LRU}} (\boldsymbol{\eta}^\star_\mathrm{RR})  + \! \sum\limits_{l \notin \mathcal{C}_m} \upsilon_l \cdot u_{m, l, \mathrm{LRU}} ( \boldsymbol{\eta}^\star_\mathrm{RR} ) \nonumber \\ 
= & \sum\limits_{l \in \mathcal{C}_m} \!\upsilon_l \!\!\!  \sum\limits_{\{k| m \in \mathcal{H}_{k,l}\}} \!\!  \rho^\mathrm{LRU}_{e(k,m)|k} \eta_{k, \mathrm{RR}}^\star - \sum\limits_{l \notin \mathcal{C}_m}  \upsilon_l  \eta_{m, \mathrm{RR}}^\star       \nonumber \\
= & \sum\limits_{l \in \mathcal{C}_m} \!\upsilon_l \!\!\!  \sum\limits_{\{k| m \in \mathcal{H}_{k,l}\}} \!\!  \rho^\mathrm{LRU}_{e(k,m)|k} \eta_{k, \mathrm{RR}}^\star -  \frac{1}{L}\sum\limits_{k \in \mathcal{H}_m }  \eta_{k, \mathrm{RR}}^\star \upsilon_{e(m, k)}      \nonumber \\
= & \sum\limits_{l \in \mathcal{C}_m} \! \upsilon_l  \!\!\! \sum\limits_{\{k| m \in \mathcal{H}_{k,l}\}} \!\!  \rho^\mathrm{LRU}_{e(k,m)|k} \eta_{k, \mathrm{RR}}^\star - \!\sum\limits_{l \in \mathcal{C}_m}  \upsilon_l  \frac{1}{L} \! \!  \sum\limits_{\{k| m \in \mathcal{H}_{k,l}\}} \!\!  \eta_{k, \mathrm{RR}}^\star       \nonumber \\
= &  \sum\limits_{l \in \mathcal{C}_m} \! \upsilon_l \!\!\!  \sum\limits_{\{k| m \in \mathcal{H}_{k,l}\}} \bigg( \rho^\mathrm{LRU}_{e(k,m)|k}  - \frac{1}{L}  \bigg)\eta_{k, \mathrm{RR}}^\star,
\end{align}
where the second step uses eq.~\eqref{e:etaSteadyLRU} and the third step uses the property in eq.~\eqref{e:etaSteadyRNProp2}. The term $\rho^\mathrm{LRU}_{e(k,m)|k}  - 1/L$ in eq.~\eqref{e:compSteadyRR_LRU} is interesting as it shows the difference between the steady states in RR and LRU. Specifically, \eqref{e:compSteadyRR_LRU} shows that, compared to RR, the steady state of LRU favors states with popular contents. 

As an example, consider the case when state $m$ caches the $L$ most popular contents. Then it follows that $\rho^\mathrm{LRU}_{e(k,m)|k} > 1/L$ in eq.~\eqref{e:compSteadyRR_LRU} for any $k$ such that $ m \in \mathcal{H}_{k}$. This is true because content $e(k,m)$ is less popular than the other $L-1$ contents in state $k$, which are also cached by state $m$ and therefore among the $L$ most popular contents. Note that the constant $1/L$ can be considered as the probability that $e(k,m)$ is the LRU content when all cached contents have exactly the same request probability. As $\rho^\mathrm{LRU}_{e(k,m)|k} > 1/L$ for any $k$ such that $ m \in \mathcal{H}_{k}$ in eq.~\eqref{e:compSteadyRR_LRU}, $u_{m, \mathrm{LRU}} (\boldsymbol{\eta}^\star_\mathrm{RR}) > 0$, which shows that the STF of the LRU at $\boldsymbol{\eta}^{\star}_\mathrm{RR}$ points towards a direction that increases the probability of caching state $m$. Similarly, it can be shown that $u_{m^\prime, \mathrm{LRU}} (\boldsymbol{\eta}^\star_\mathrm{RR}) < 0$ if $m^\prime$ caches the least popular contents.       

The above difference between the steady states of the RR and LRU roots from the difference in the information exploited in the two schemes. Unlike RR, which exploits no information and treats each cached content indifferently in every single replacement, the LRU exploits historical request information, which reflects the content popularity. As a result, LRU can converge to a steady state that caches popular contents with larger probabilities.

\section{Discussions}\label{s:Dis}

In this section, we discuss the benefits of using the proposed STF to analyze replacement schemes in practice. First, we use an example to show how the STF can characterize the property of the stationary states. Then, we use another example to show how the STF can be used to compare the convergence rate of replacement schemes.

\subsection{On the Steady State}

Given two replacement schemes (or the same replacement scheme with different parameters), can we tell more about their steady states besides the cache hit probability?

\begin{figure}[t]
		\centering 
		\subfloat[Small $\|\mathbf{u}_l(\boldsymbol{\eta}^\star)\|$.] %
		{\includegraphics[angle=0,width=0.24\textwidth]{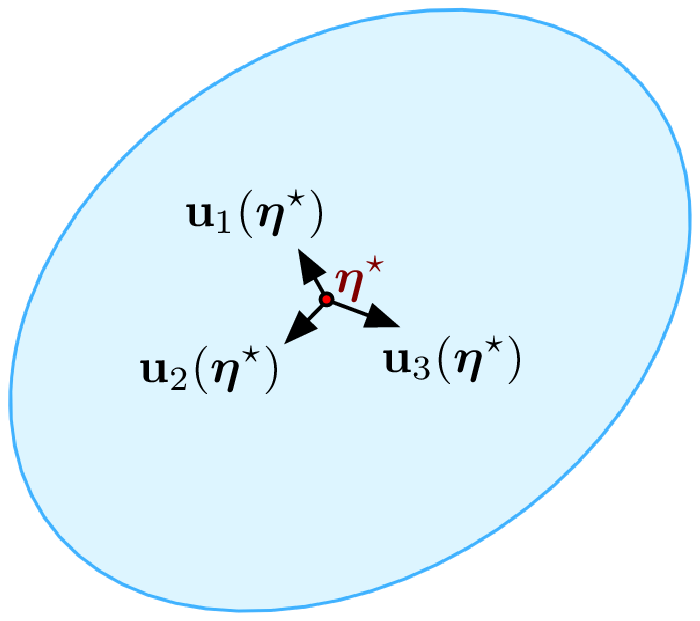}\label{f:SOsc1}} 
		\hspace{-2mm} 
		\subfloat[Large $\|\mathbf{u}_l(\tilde{\boldsymbol{\eta}}^\star)\|$.]
		{\includegraphics[angle=0,width=0.24\textwidth]{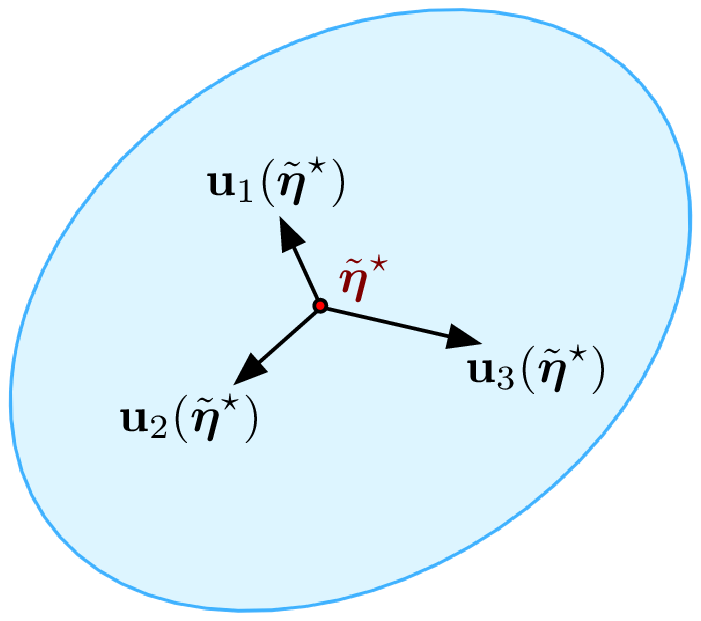}\label{f:SOsc2}}
		\caption{Illustration of decomposing the STF at the steady state.}\label{f:SOsc} \vspace{-2mm}
\end{figure}

At the steady state, the overall STF must be equal to $\mathbf{0}$ regardless of the replacement scheme. However, this does not mean that no replacement happens after the steady state is achieved. Instead, contents can still be evicted from or accepted into the cache, while the probabilities of the two events must be equal for any content at the steady state. Therefore, it is not difficult to see that, there can be more frequent replacements at the steady state of one replacement scheme than that of another. This frequency of replacement at a steady state can be analyzed by decomposing the STF into content-specific STF using eq.~\eqref{e:u_Decomp}, as illustrated in Fig.~\ref{f:SOsc}. In the illustrated cases, we assume the same content request probabilities, while the content-specific STFs in Fig.~\ref{f:SOsc1} have much smaller norms than those in Fig.~\ref{f:SOsc2}. Correspondingly, there can be less frequent replacements at the steady state $\boldsymbol{\eta}^\star$ in Fig.~\ref{f:SOsc1} than at the steady state $\tilde{\boldsymbol{\eta}}^\star$ in Fig.~\ref{f:SOsc2}.

In the case when each replacement incurs a cost or when  cache wear-out is a concern, characterizing the frequency of replacement can be of interest. Based on the above discussion, the weighted sum of the norm of content-specific STF can be used as a metric for comparing the frequency of content replacement at the steady state of different replacement schemes. For example, a metric can be calculated as follows: 
\begin{align}  
M(\boldsymbol{\eta}^\star) = \sum\limits_{l \in \mathcal{C}} \upsilon_l \|\mathbf{u}_l(\boldsymbol{\eta}^\star)\|, 
\end{align}
where the weights are the content request probabilities.

\subsection{On the Convergence to the Steady State}

We mentioned the rate of convergence and its relation with the second largest eigenvalue of the transition probability matrix $\boldsymbol{\Theta}$ in Section~\ref{s:Analysis}. Since STF is a derivative of state transition probability matrix, it does not provide a new characterization of the rate of convergence in theory. However, we could use STF to develop a metric for comparing the convergence rate of different replacement schemes in practice.     

\begin{figure}[t]
	\centering 
	\includegraphics[angle=0,width=0.30\textwidth]{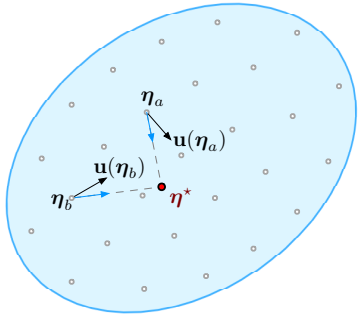}
	\caption{Illustration of sampling the STF for characterizing the convergence rate.}\label{f:ConvergeSTF} 
	\vspace{-2mm}
\end{figure}

For example, we can generate sample points in the state transition region, as illustrated using hollow circles in Fig.~\ref{f:ConvergeSTF}. Hypothetically, if the STF at every point of the state transition region points toward the steady state $\boldsymbol{\eta}^\star$, then the rate of convergence is determined by the strength (norm) of the STF. In practice, the STF at the sample points generally does not point straight toward the steady state. Nevertheless, we can project the STF at a sample point onto the connection line between that sample point and the steady state. This is illustrated with two example sample points, i.e., $\boldsymbol{\eta}_a$ and $\boldsymbol{\eta}_b$, in Fig.~\ref{f:ConvergeSTF}. In this figure, the solid circle filled with red represents the steady state $\boldsymbol{\eta}^\star$. The black arrows at sample points $\boldsymbol{\eta}_a$ and $\boldsymbol{\eta}_b$ represent the STF $\mathbf{u}(\boldsymbol{\eta}_a)$ and $\mathbf{u}({\boldsymbol{\eta}}_b)$, respectively. The two dashed lines connect $\boldsymbol{\eta}_a$ and $\boldsymbol{\eta}_b$ with the steady state $\boldsymbol{\eta}^\star$. The two blue arrows on the dashed lines represent the projection of $\mathbf{u}(\boldsymbol{\eta}_a)$ and $\mathbf{u}({\boldsymbol{\eta}}_b)$, respectively. The norm of the projection, aggregated over all sample points, can provide a metric for characterizing the rate of convergence of replacement schemes. The accuracy of this approach depends on the number and locations of the chosen sample points.


\section{Numerical Results}\label{s:Simu}

The numerical examples are organized into three sections. The first section demonstrates STFs obtained from analysis. The second section demonstrates STFs obtained from simulations and compare it with the analytical results. The third section demonstrates and compares the CCP and cache hit probability of the considered schemes to reveal the impact of different STFs.  

\subsection{STF - Analytical}

In this section, the analytical STFs of RR, LP, TLP, and LRU are demonstrated. In general, STF can be of high dimensions. We limit most of our demonstration to the case of three dimensions, as three-dimensional fields can be very well visualized and illustrated. A three-dimensional subspace in a high-dimensional STF is also illustrated.

\begin{figure}[t]
	\centering 
	\subfloat[{STF of RR, $\phi = 0.45, \boldsymbol{\upsilon} = [0.5, 0.29, 0.21]^\mathrm{T}$}.] %
	{\includegraphics[angle=0,width=0.39\textwidth]{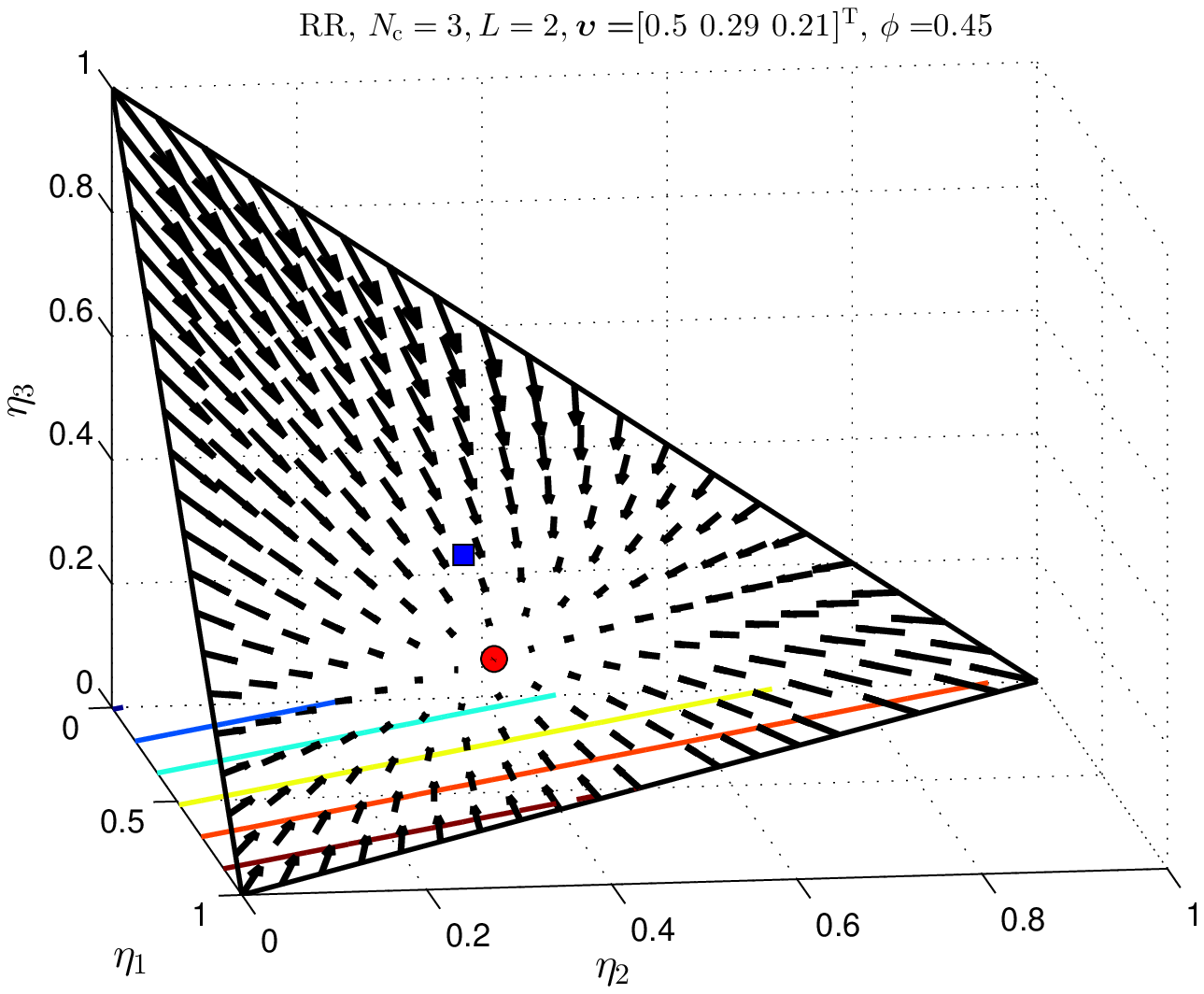}\label{f:RR3D1}} 
	\hspace{-0mm} 
	\subfloat[STF of RR, $N_\mathrm{c} = 30, L= 3$, in a 3-D subspace.]
	{\includegraphics[angle=0,width=0.39\textwidth]{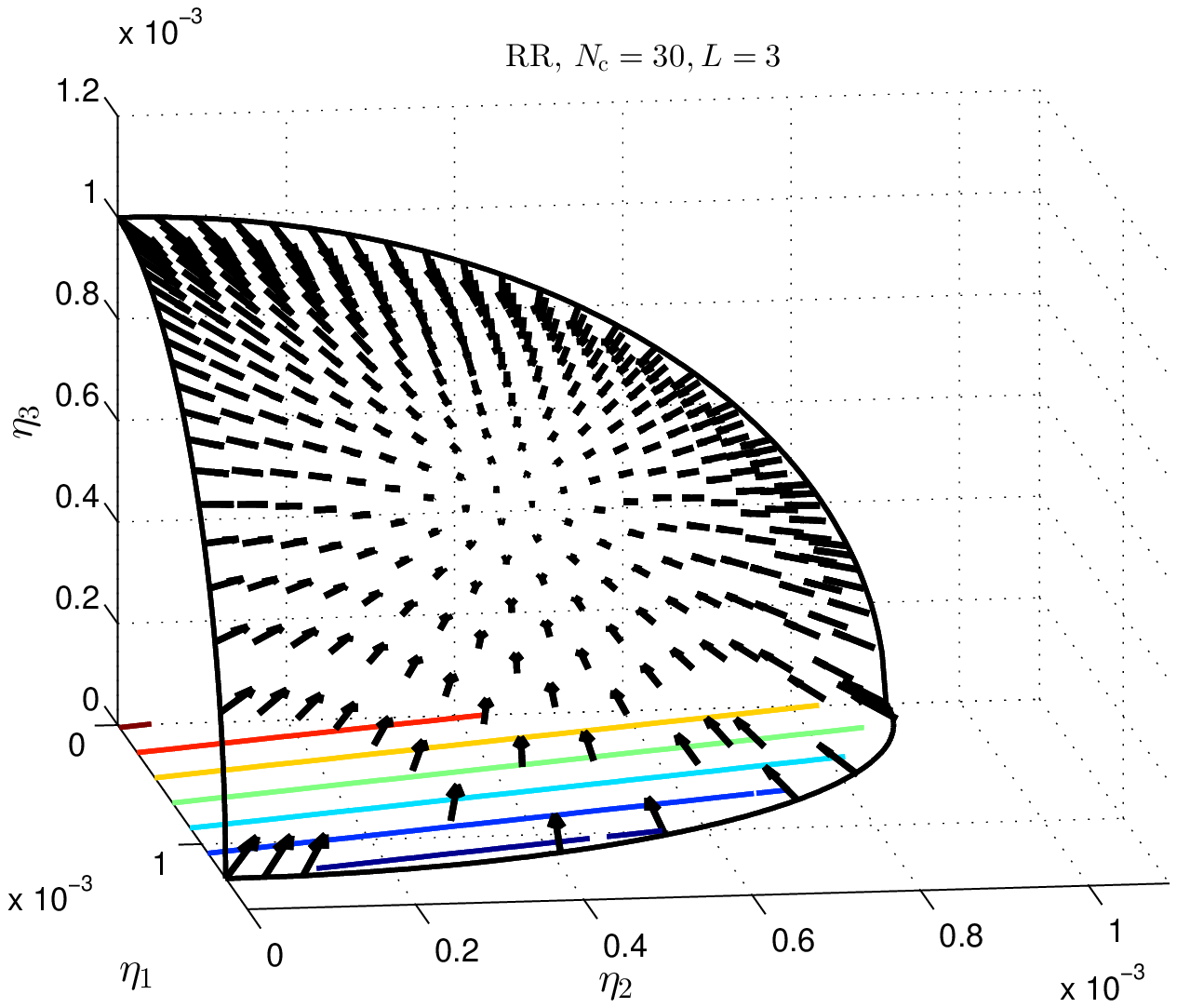}\label{f:RR3DHigh}}
	\caption{STF of RR in 3-D.}\label{f:field_RR3andHigher} \vspace{-2mm}
\end{figure}

Fig.~\ref{f:RR3D1} demonstrates a three-dimensional STF of RR. In this figure, $N_\mathrm{c}= 3$, $L=2$, and therefore there are only three cache states (i.e., $\mathcal{C}_1 = \{1, 2\}, \mathcal{C}_2 = \{1, 3\}, \mathcal{C}_3 = \{2, 3\}$). The $x$, $y$, and $z$ axes correspond to the SCP for the cache states 1, 2, and 3, respectively. The triangular area is the state transition domain $\mathcal{D}$, the square marker represents the center of the triangle, and the circle represents the steady-state SCP $\boldsymbol{\eta}^{\star}$ in this example. The STF at a point in $\mathcal{D}$ is represented by an arrow originating from that point, while the strength and direction of the STF are shown by the length of the arrow and the direction of the arrowhead, respectively. The straight lines in the x-y plane show the contour of the cache hit probability for the SCP.

Fig.~\ref{f:RR3DHigh} demonstrates part of a high-dimensional STF over the surface of an ellipsoid in a three-dimensional subspace. In this example, $N_\mathrm{c}= 30$, $L=3$, and there are 4060 cache states. Three mutually-neighbor cache states are selected, corresponding to the three-dimensional subspace in the figure. The STF over the surface of an ellipsoid in this subspace is demonstrated as an example. The $x$, $y$, and $z$ axes correspond to the SCP for the three selected cache states. Unlike the case in Fig.~\ref{f:RR3D1}, the SCPs in ~\ref{f:RR3DHigh} are small and do not sum up to 1 since there are many other states. Fig.~\ref{f:RR3DHigh} serves as an example of  high-dimensional STF.

\begin{figure}[t]
	\centering 
		\subfloat[{STF of RR, $\phi = 0.2, \boldsymbol{\upsilon} = [0.5, 0.29, 0.21]^\mathrm{T}$}.]
	{\includegraphics[angle=0,width=0.38\textwidth]{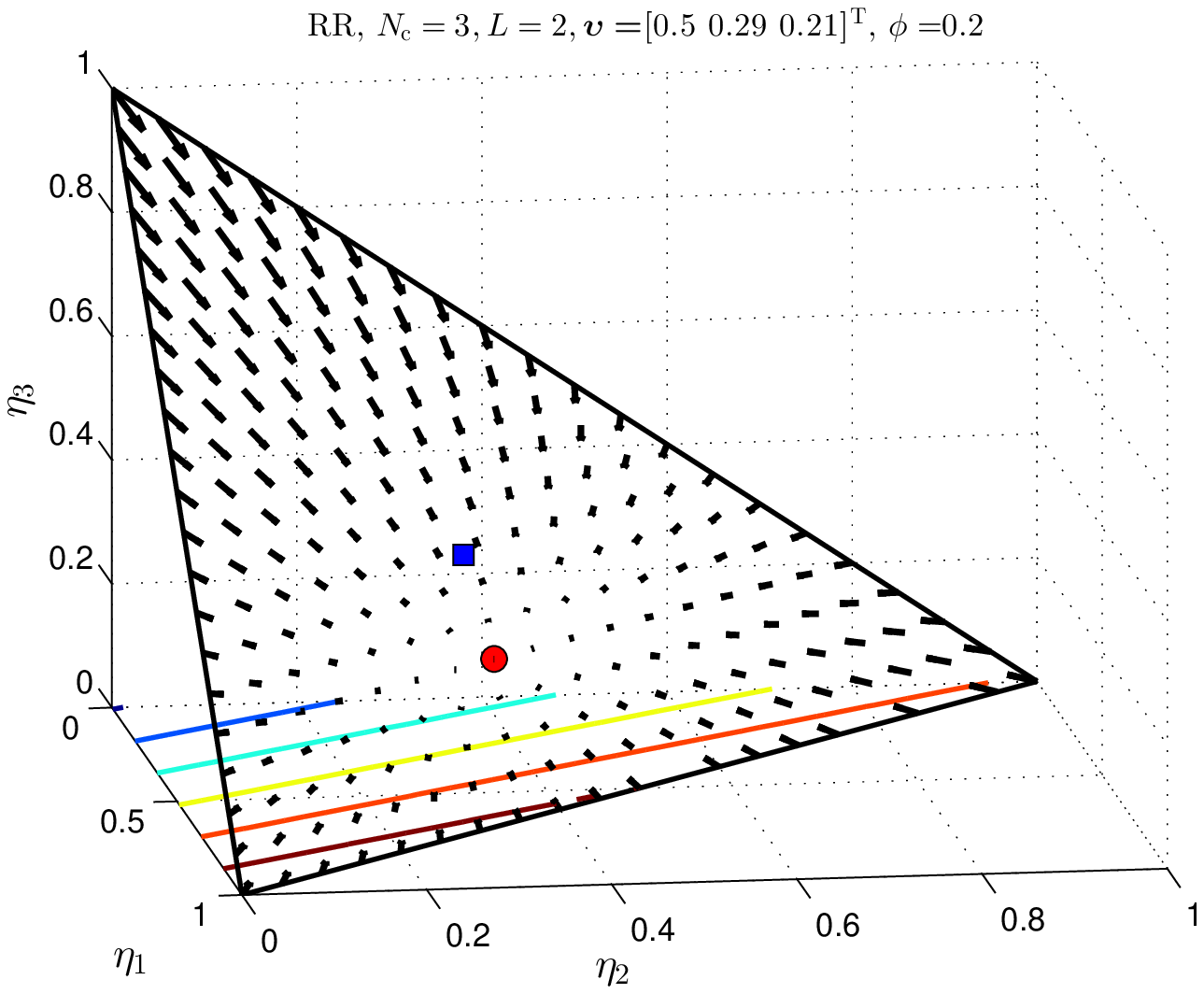}\label{f:RR3D2}}
		\hspace{1mm} 
		\subfloat[{STF of RR, $\phi = 0.45, \boldsymbol{\upsilon} \!=\! [0.55, 0.35, 0.1]^\mathrm{T}$}.]
	{\includegraphics[angle=0,width=0.38\textwidth]{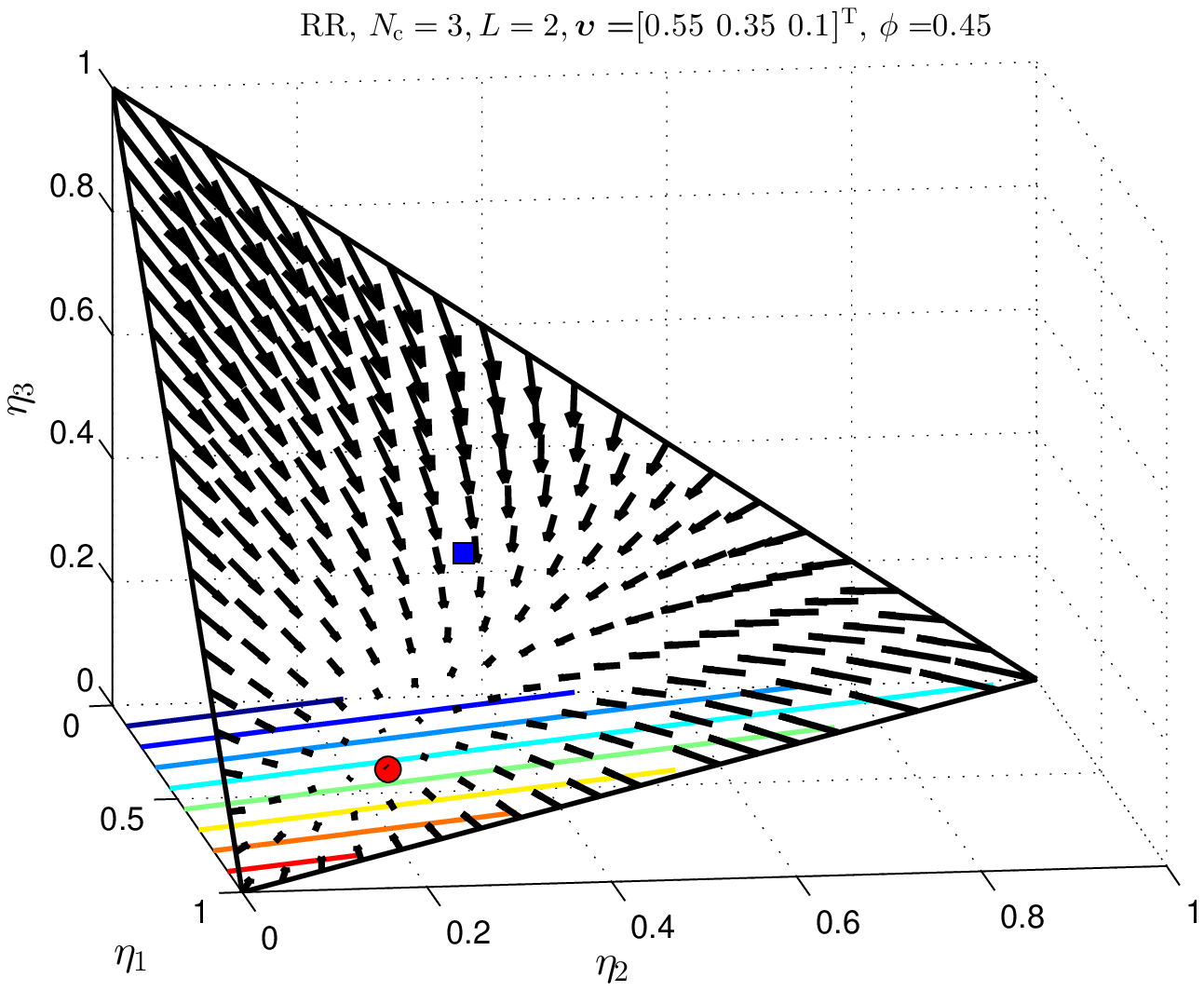}\label{f:RR3D3}}
	\caption{The impact of $\boldsymbol{\upsilon}$ and $\phi$ on the STF of RR.}\label{f:fieldRR3D} 
\vspace{-2mm}
\end{figure}

Fig.~\ref{f:fieldRR3D} demonstrates the impact of the content popularity $\boldsymbol{\upsilon}$ and the parameter $\phi$ on the STF of RR. Fig.~\ref{f:RR3D2} shows the STF under the same settings as in Fig.~\ref{f:RR3D1} except that $\phi$ is decreased from 0.45 to 0.2. Two observations can be made by comparing Fig.~\ref{f:RR3D2} with Fig.~\ref{f:RR3D1}. First, the steady-state SCP in both cases are identical, which confirms Theorem~\ref{t:PropertyRRsteady}.  Second, the strength of STF at any given point in Fig.~\ref{f:RR3D2} is weaker as compared to that in Fig.~\ref{f:RR3D1}, which implies a longer mixing time. Fig.~\ref{f:RR3D3} shows the STF under the same settings as in Fig.~\ref{f:RR3D1} except a change in the content popularity $\boldsymbol{\upsilon}$. It can be seen from this figure that the steady state also changes following the change in content popularity. Comparing Fig.~\ref{f:RR3D3} with Fig.~\ref{f:RR3D1}, the impact of content popularity on the STF can be observed. 

\begin{figure}[t]
	\centering 
	\subfloat[{STF of LP, $\alpha = 0.9$}.]
	{\includegraphics[angle=0,width=0.38\textwidth]{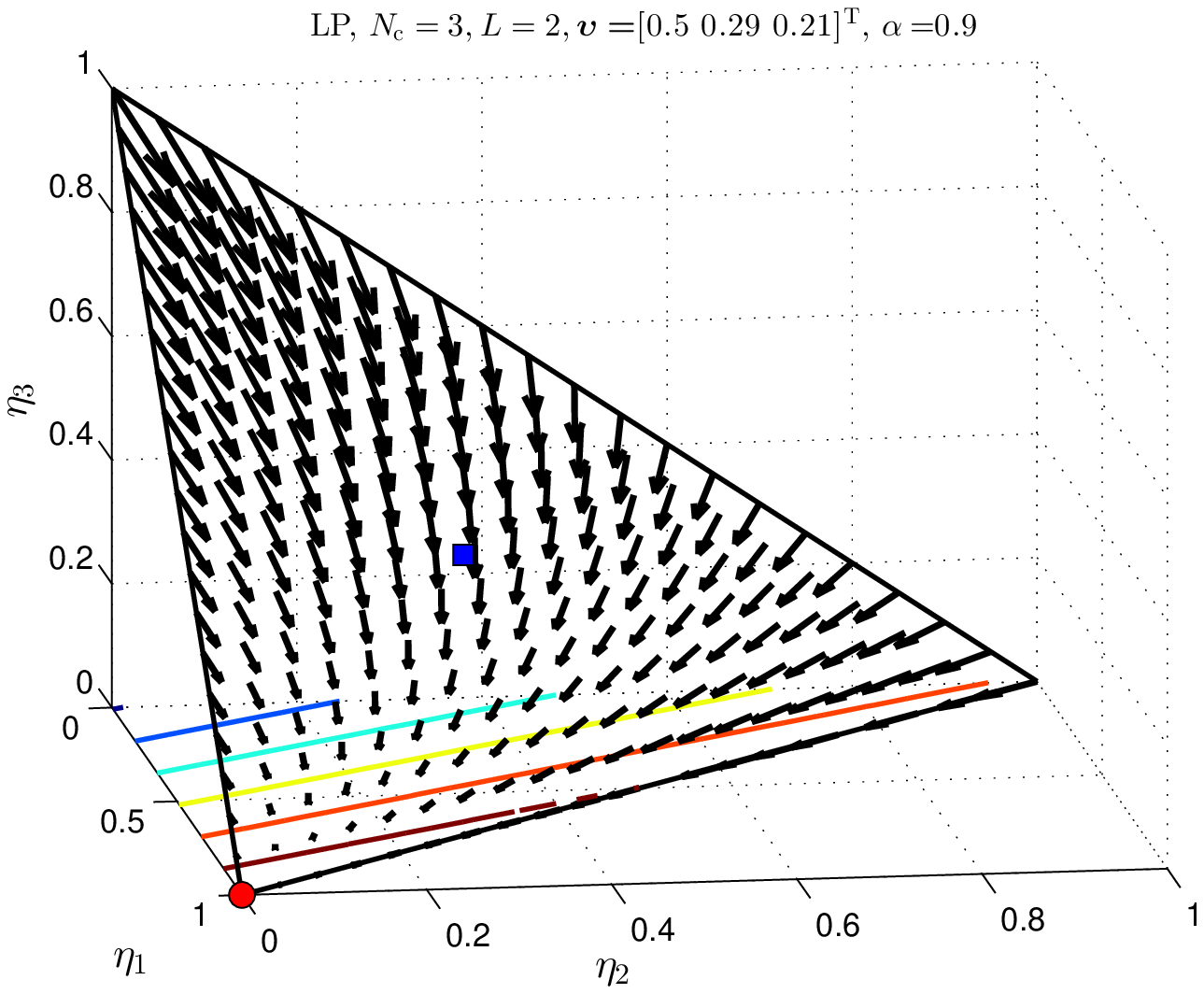}\label{f:LES3D}}
	\hspace{1mm} 
	\subfloat[{STF of TLP-A}]
	{\includegraphics[angle=0,width=0.38\textwidth]{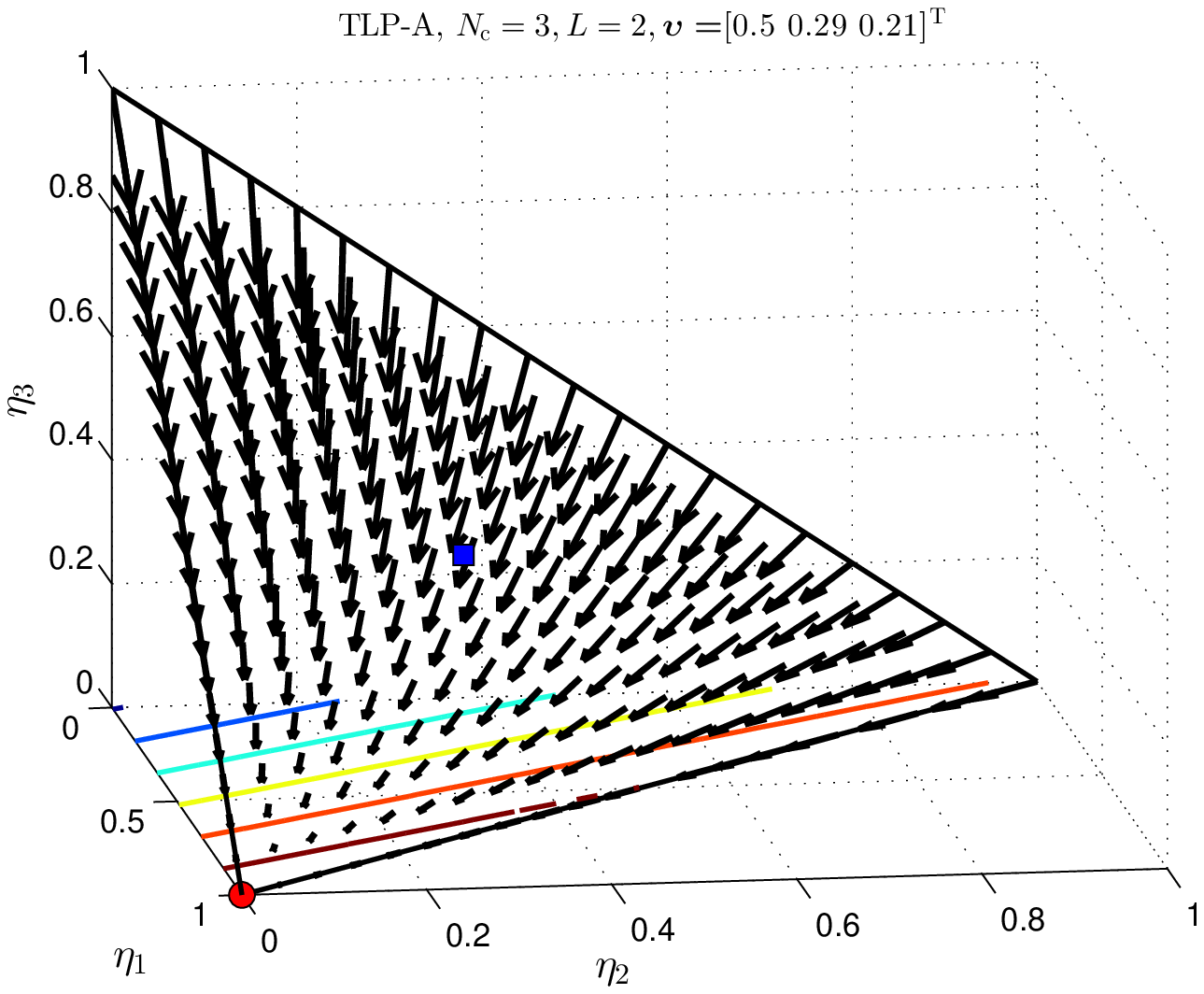}\label{f:LEA3D}}
	\hspace{-1mm} 
	\subfloat[STF of LRU]
	{\includegraphics[angle=0,width=0.38\textwidth]{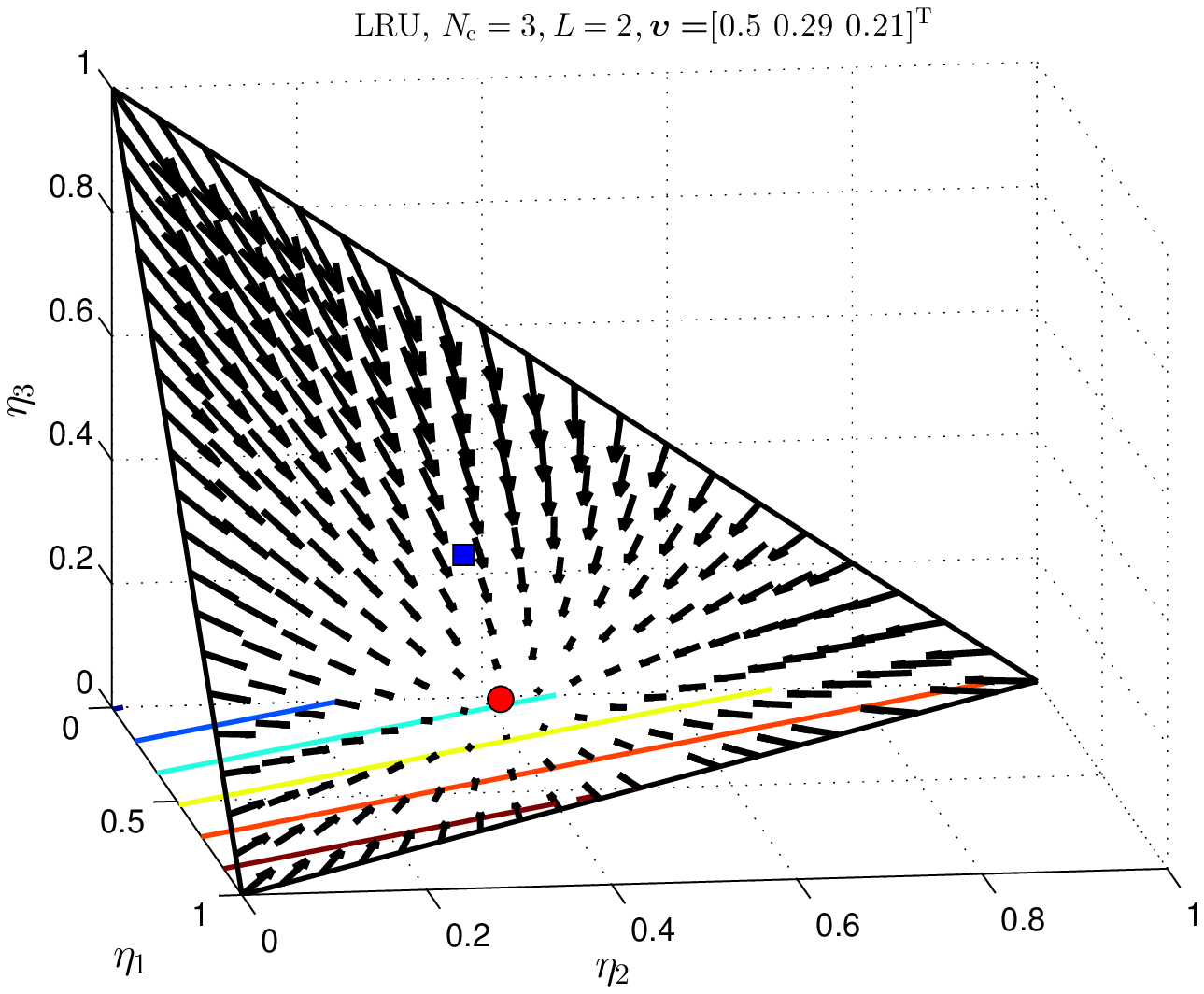}\label{f:LRU3D}}
	\caption{The STF of LP, TLP, and LRU in 3-D.}\label{f:fieldOthers3D} \vspace{-2mm}
\end{figure}

Fig.~\ref{f:fieldOthers3D} demonstrates three-dimensional STFs of LP, TLP, and LRU. In all three plots in Fig.~\ref{f:fieldOthers3D}, $\boldsymbol{\upsilon}$ is set to $[0.5, 0.29, 0.21]^\mathrm{T}$. In Figs.~\ref{f:LES3D}~and~\ref{f:LEA3D}, the steady state is the vertex of the triangle with the highest cache hit probability. The difference is that the STF in Fig.~\ref{f:LES3D} lead to a `curvy' path towards the steady state in Fig.~\ref{f:LES3D} while the curvature of paths in Fig.~\ref{f:LEA3D} is much smaller. This reflects the fact that TLP makes replacements along the path which increases the cache hit probability most rapidly, bearing a certain resemblance to the `steepest ascent' in gradient ascent. Fig.~\ref{f:LRU3D} appears similar to Fig.~\ref{f:RR3D1}. However, it can be observed that, compared to RR, the steady state of LRU assigns a larger caching probability to states with more popular contents. This is consistent with eq.~\eqref{e:compSteadyRR_LRU} and the fact that RR exploits no historical information while making cache replacements.

\subsection{STF - Numerical} 

In this section, we demonstrate, using RR and LRU as examples, STFs obtained from simulations and compare them with the analytical STF from the preceding section.

\begin{figure}[t]
	\centering 
	\subfloat[{STF of RR from simulation, $M = 100$, $R = 100$}.]
	{\includegraphics[angle=0,width=0.38\textwidth]{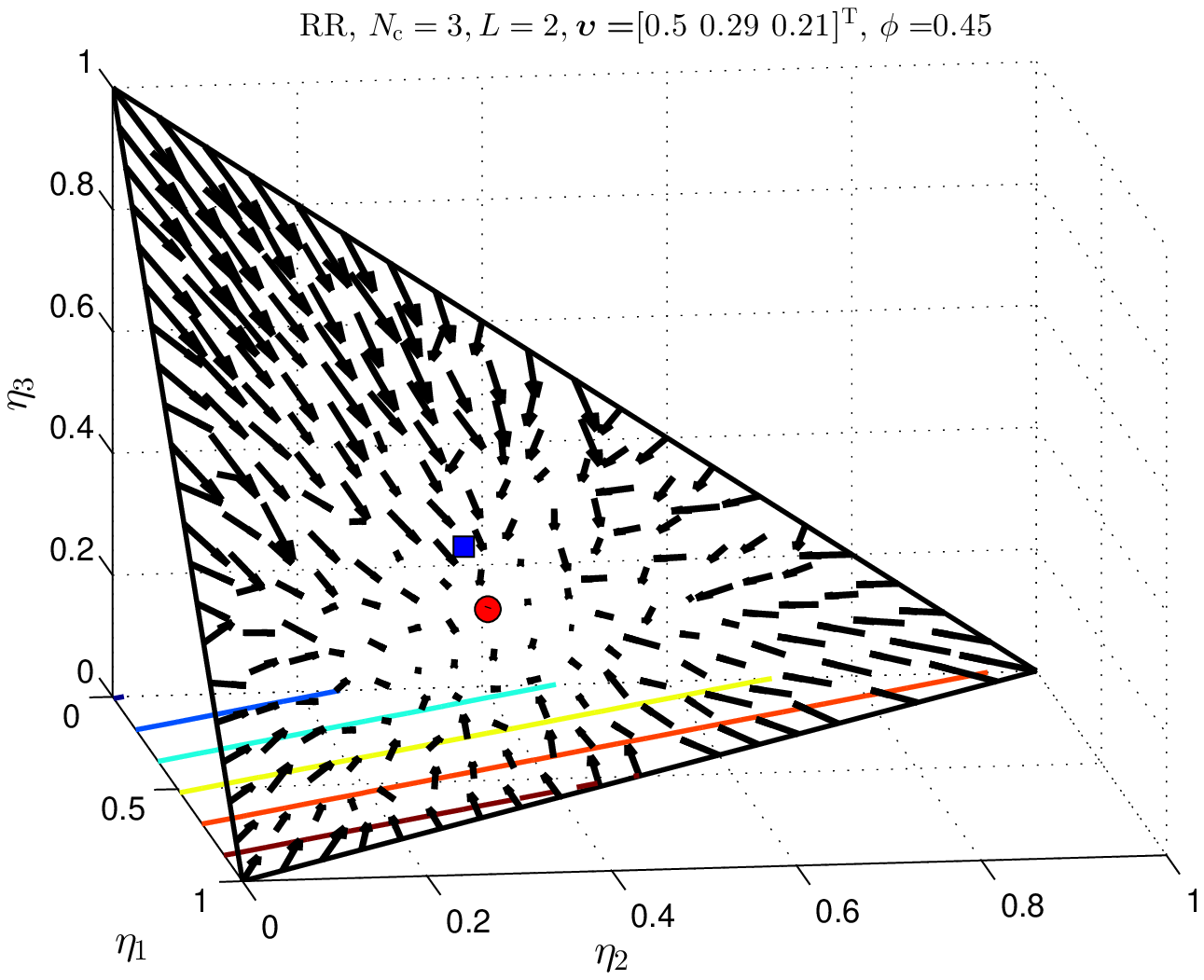}\label{f:RR5Simu}}
	\hspace{-1mm} 
	\subfloat[{{STF of RR from simulation, $M = 1000$, $R = 1000$.}}]
	{\includegraphics[angle=0,width=0.38\textwidth]{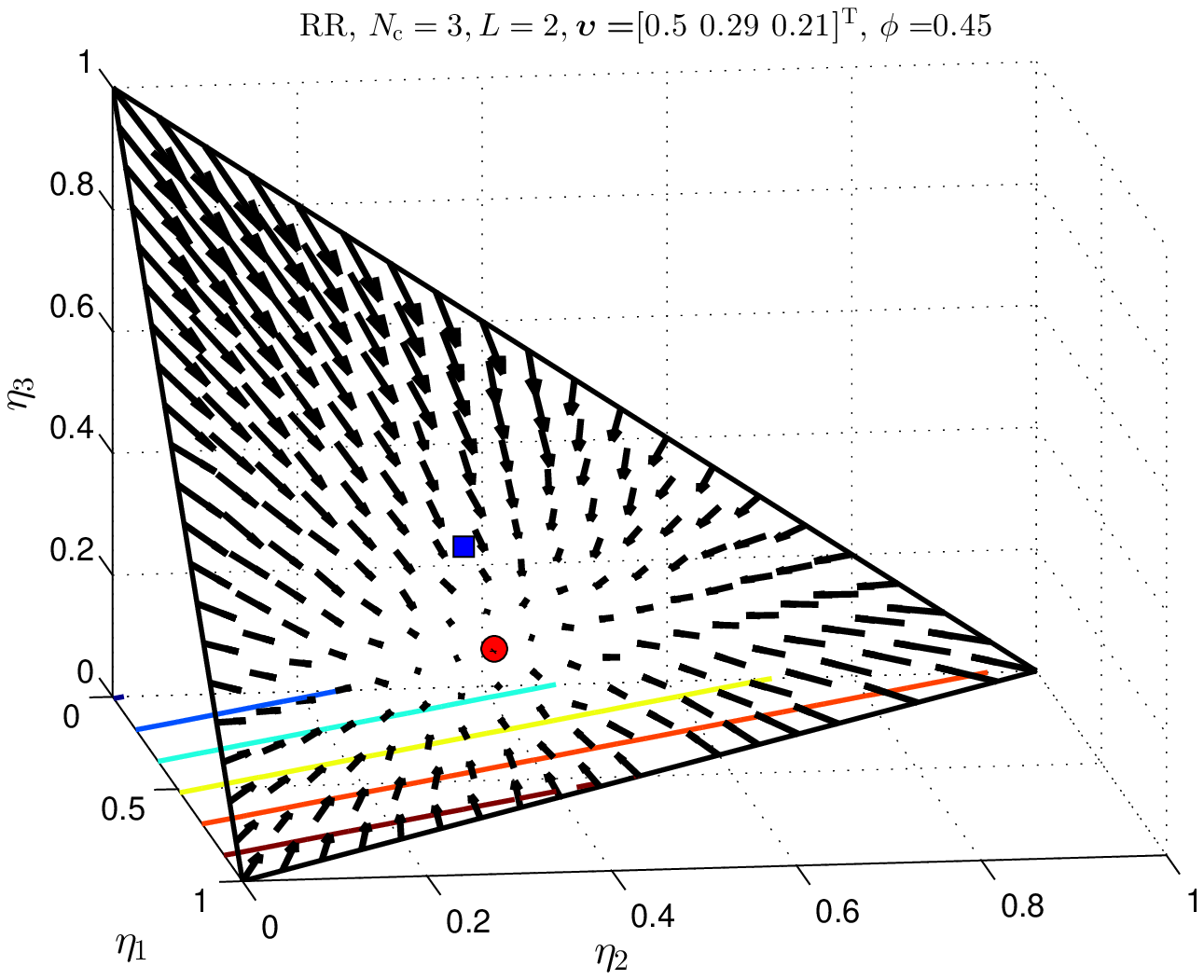}\label{f:RR6Simu}}
	\hspace{-1mm} 
	\caption{The STF of RR from simulations.}\label{f:RRSimu} \vspace{-2mm}
\end{figure}

Fig.~\ref{f:RRSimu} shows the STF of RR generated from simulations. The settings on $\phi$ and $\boldsymbol{\upsilon}$ in Fig.~\ref{f:RRSimu} are exactly the same as those in Fig.~\ref{f:RR3D1}. For each point in the STF, $M$ realizations of states are generated based on the corresponding SCP. For each realization, $R$ content requests are generated based on the content popularity. Each data point (i.e., each arrow) in Figs.~\ref{f:RR5Simu}~and~\ref{f:RR6Simu} is obtained from averaging the state transitions following the $M\times R$ requests. In Fig.~\ref{f:RR5Simu}, $M$ and $R$ are both set to 100. It can be seen that the STF is not accurate, especially in the area close to the steady state, due to insufficient samples. In addition, the arrows point to a steady state slightly deviated from the true steady state in Fig.~\ref{f:RR3D1}. In Fig.~\ref{f:RR6Simu}, $M$ and $R$ are both increased to 1000. It can be seen that the resulting STF generated based on simulation in Fig.~\ref{f:RR6Simu} becomes an exact match for the analytical STF in Fig.~\ref{f:RR3D1}.

\begin{figure}[t]
	\centering 
	\subfloat[{STF of LRU from simulation, $R = 500$}.]
	{\includegraphics[angle=0,width=0.38\textwidth]{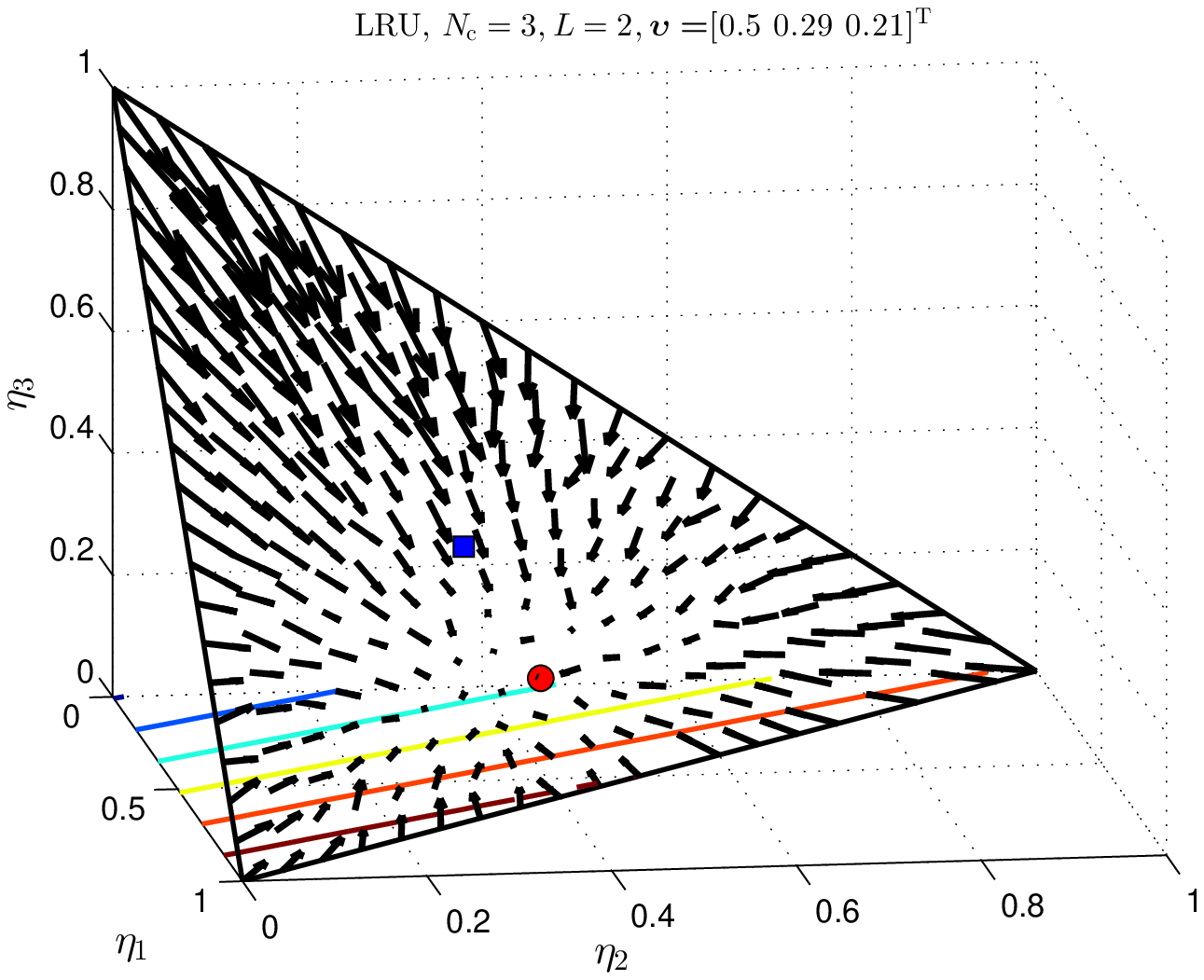}\label{f:LRU1Simu}}
	\hspace{-1mm} 
	\subfloat[{STF of LRU from simulation, $R = 10000$.}]
	{\includegraphics[angle=0,width=0.38\textwidth]{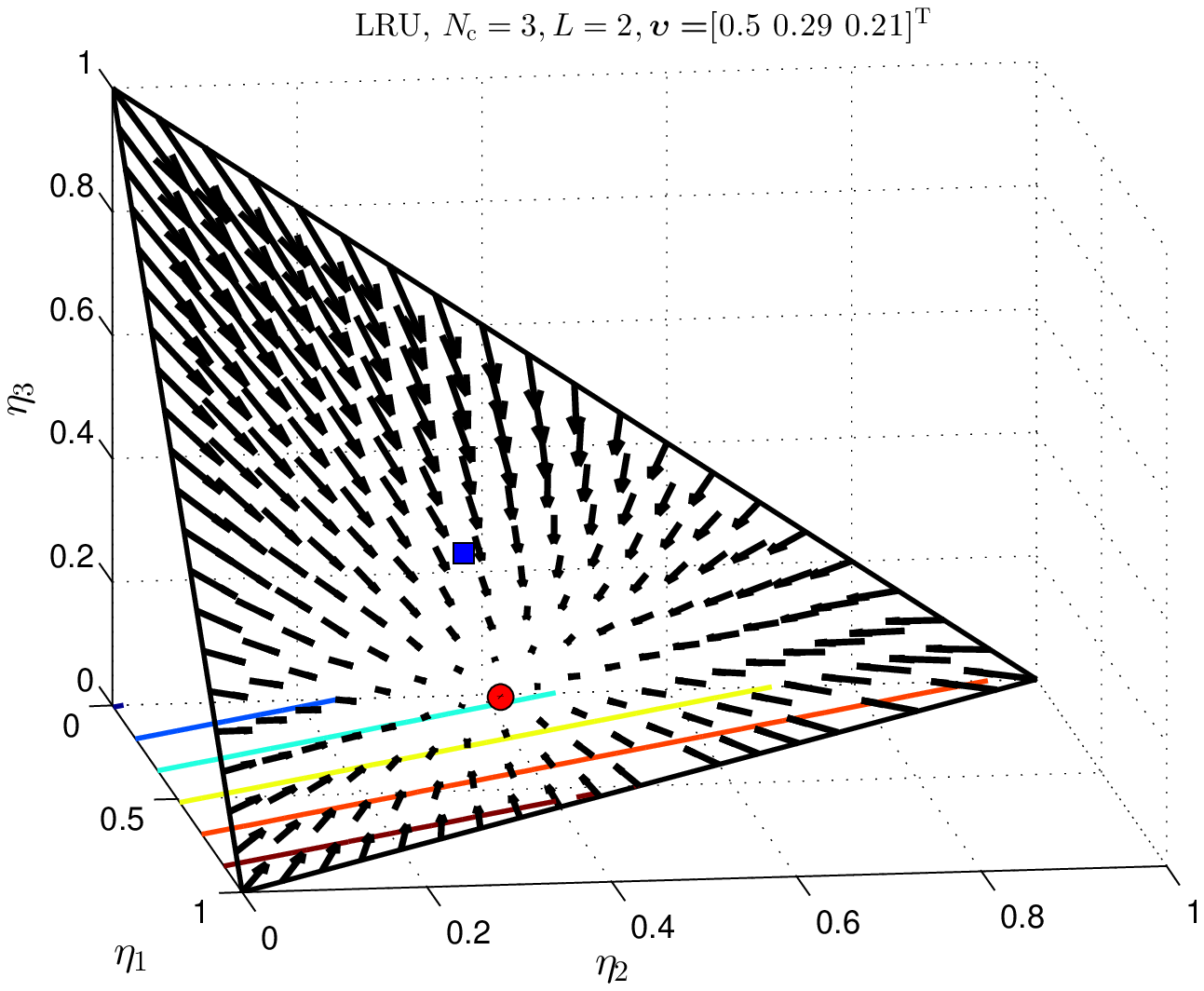}\label{f:LRU2Simu}}
	\hspace{-1mm} 
	\caption{The STF of LRU from simulations.}\label{f:LRUSimu} \vspace{-2mm}
\end{figure}

Fig.~\ref{f:LRUSimu} shows the STF of LRU generated from simulations. The settings on $\boldsymbol{\upsilon}$ in Fig.~\ref{f:LRUSimu} is exactly the same as that in Fig.~\ref{f:LRU3D}. Since LRU depends on request history, the simulation method used for Fig.~\ref{f:RRSimu} based on randomly generated states cannot be applied. Instead, for each point in the STF, $R$ content requests are generated based on the content popularity. The STF is generated based on the state transitions following the $R$ requests. Figs.~\ref{f:LRU1Simu}~and~\ref{f:LRU2Simu} demonstrate a similar result as that from Figs.~\ref{f:RR5Simu}~and~\ref{f:RR6Simu}: the STF from simulations can deviate from the analytical STF when the number of samples is small, while the two become an almost exact match when the sampled number of requests is sufficiently large.

\subsection{Instantaneous CCP}

\begin{figure*}[t]
	\centering 
	\subfloat[Instantaneous CCP - RR.]
	{\includegraphics[angle=0,width=0.410\textwidth]{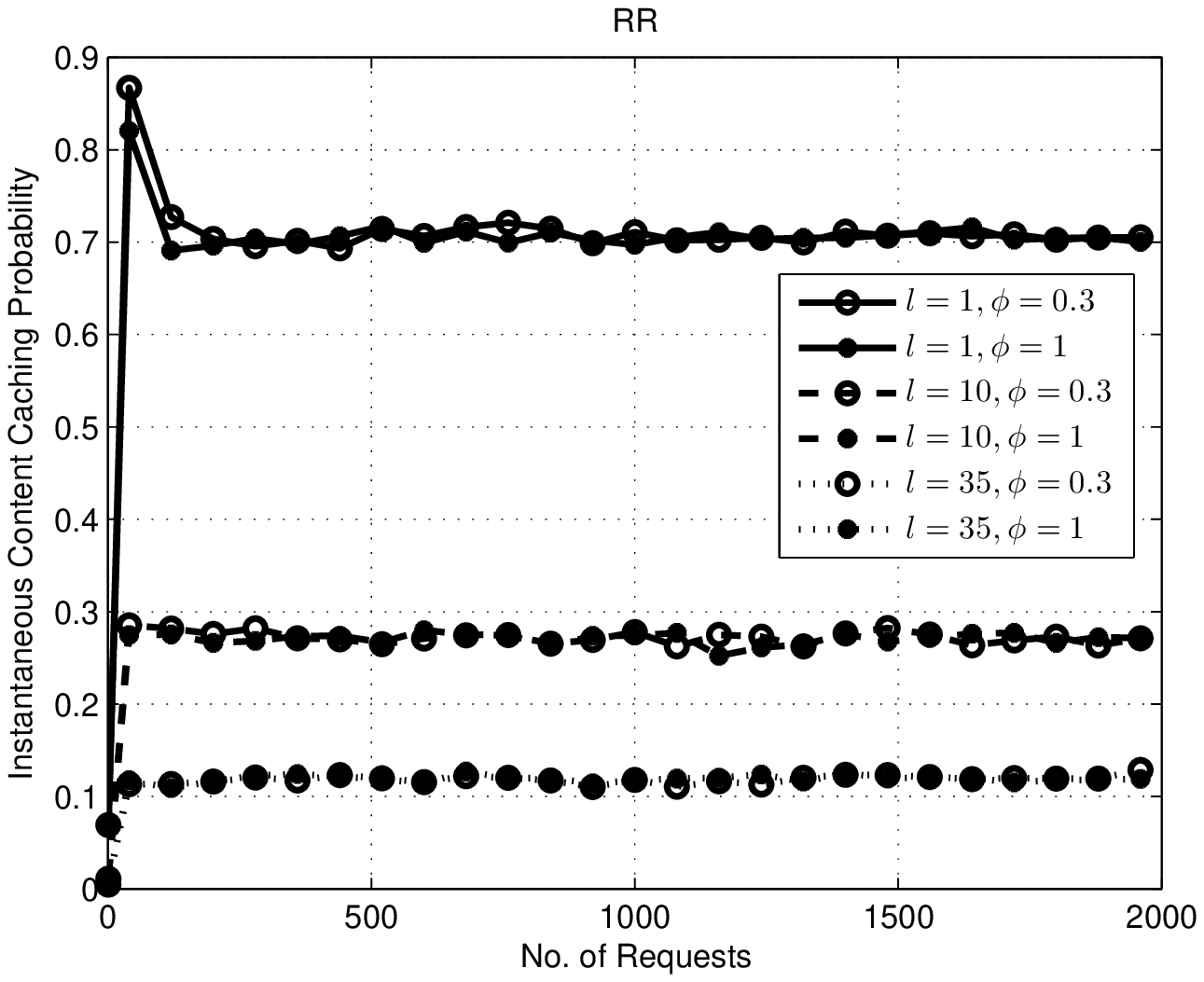}\label{f:InstCacheRR}}
	\hspace{-1mm}
	\subfloat[Instantaneous CCP - LP.]
	{\includegraphics[angle=0,width=0.410\textwidth]{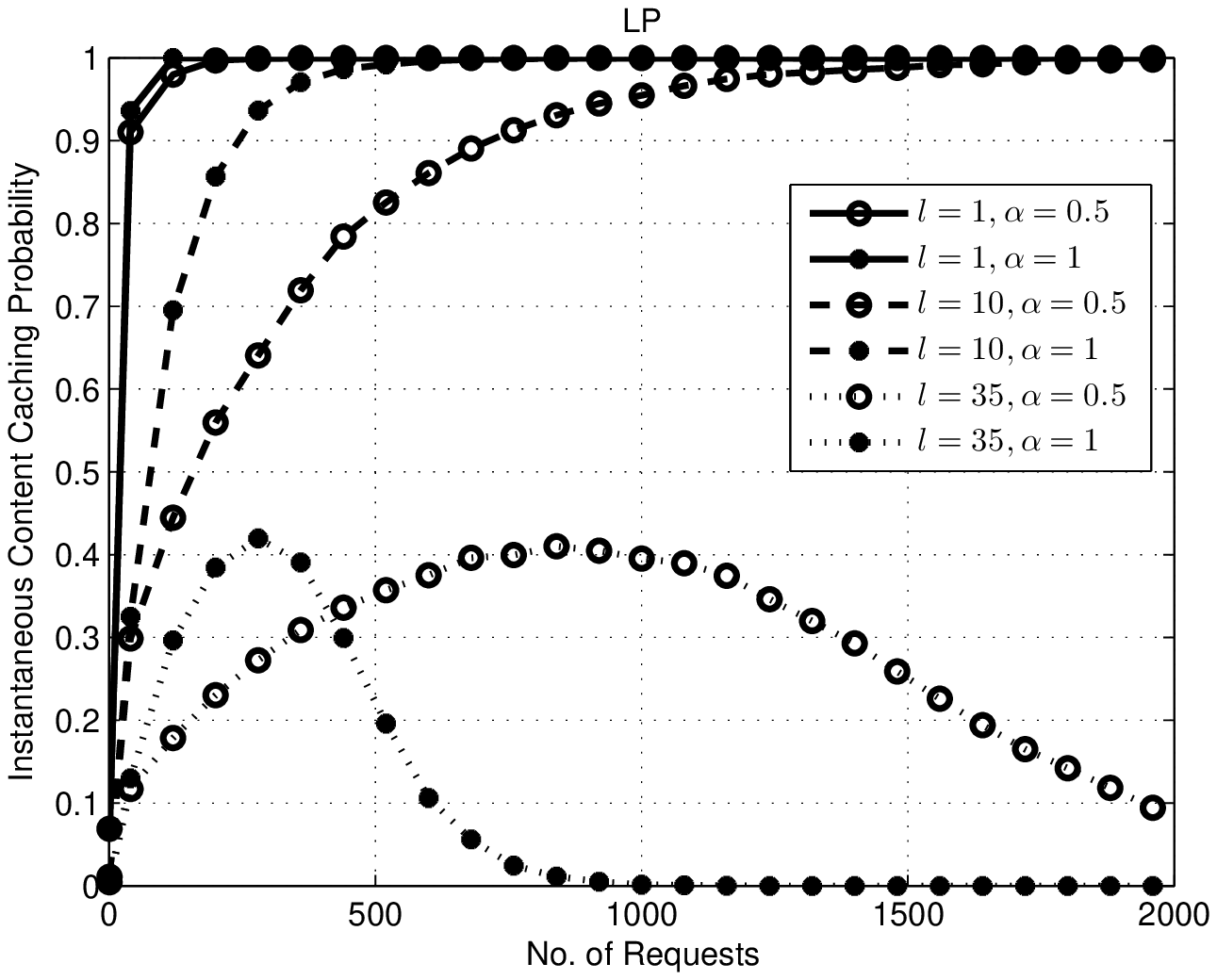}\label{f:InstCacheLES}}
	\hspace{1mm}
	\subfloat[Instantaneous CCP - TLP.]
	{\includegraphics[angle=0,width=0.410\textwidth]{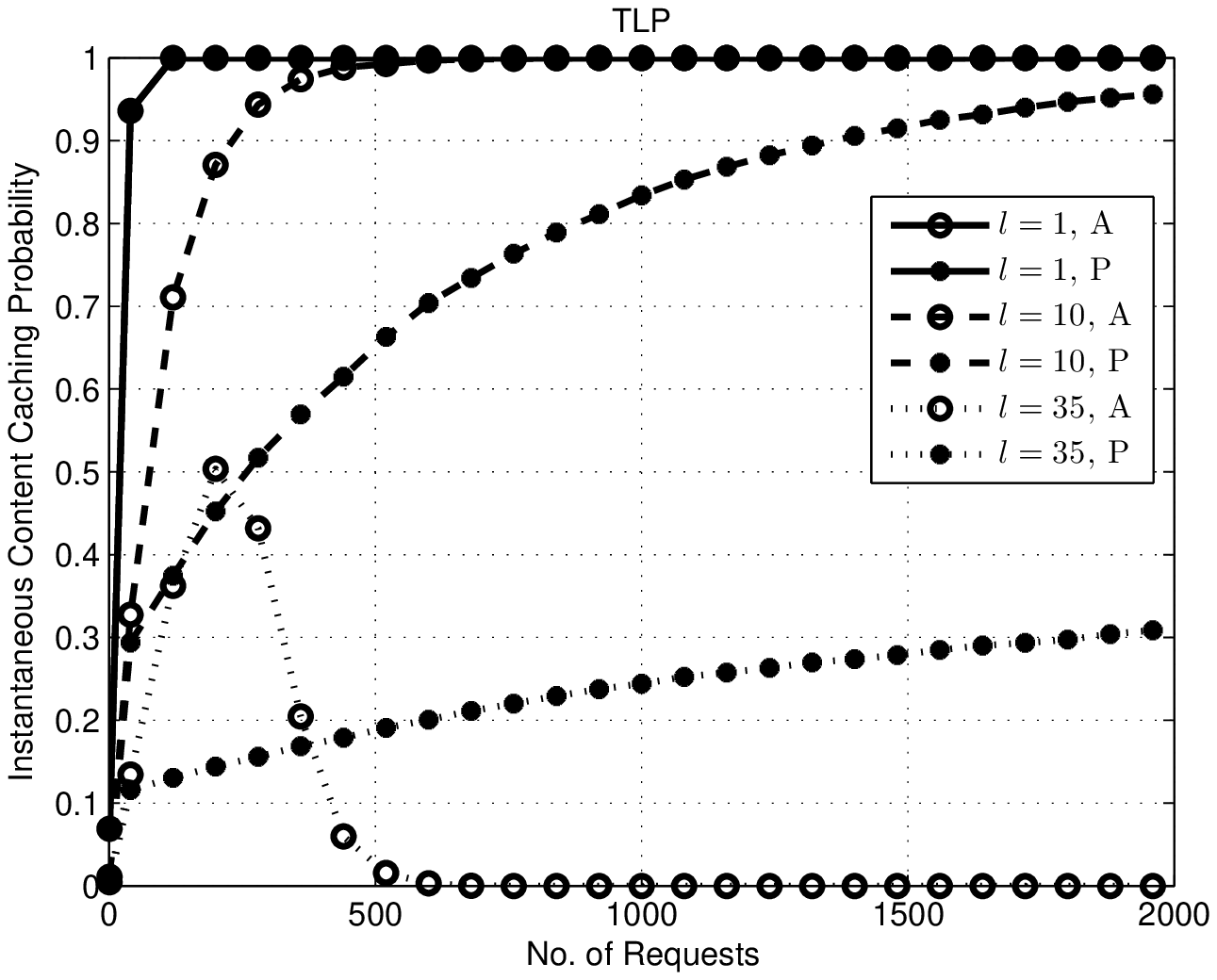}\label{f:InstCacheLEA}}
	\hspace{-1mm} 
	\subfloat[Instantaneous CCP - LRU.]
	{\includegraphics[angle=0,width=0.410\textwidth]{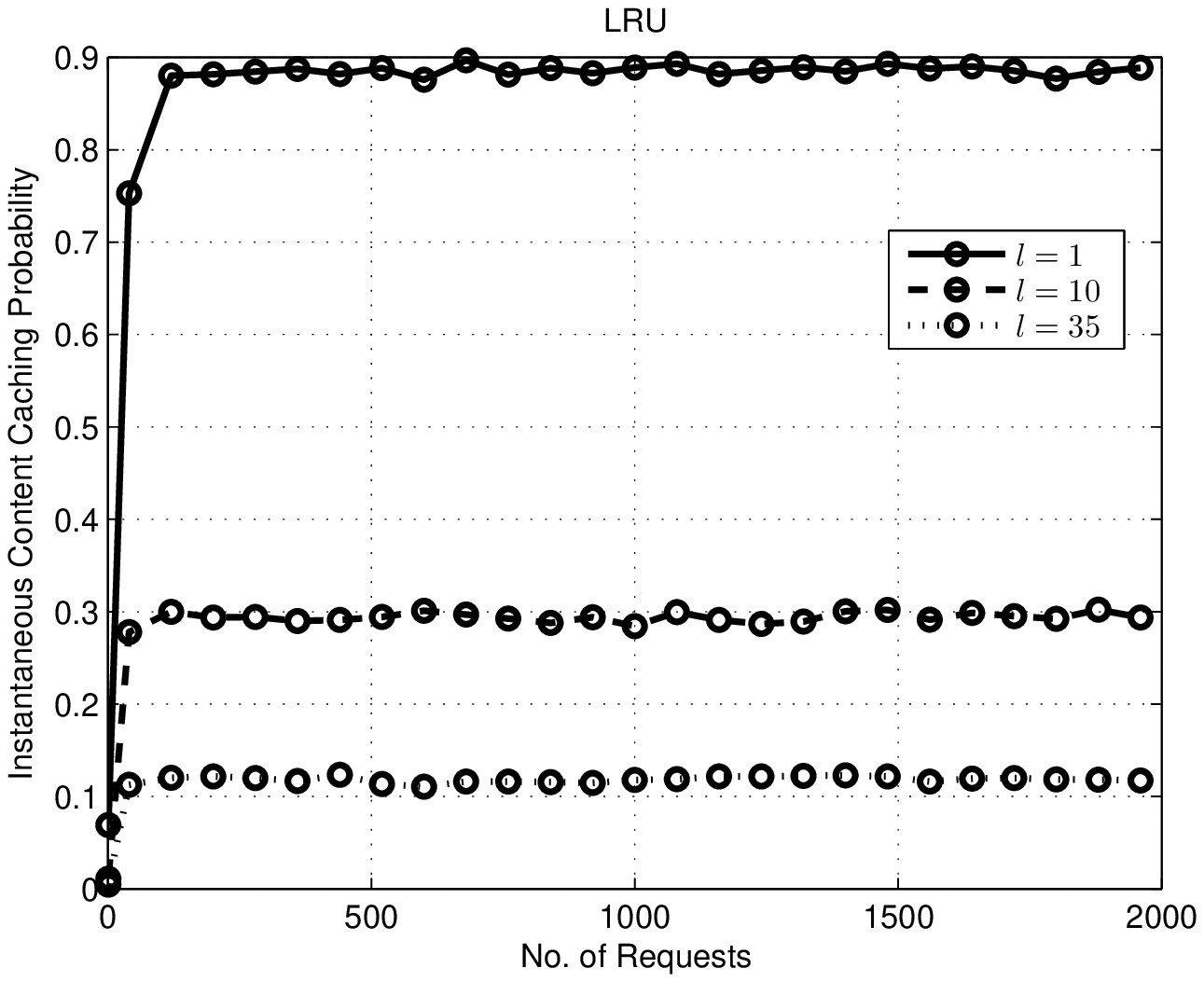}\label{f:InstCacheLRU}}
	\caption{Demonstration of instantaneous CCP of the replacement schemes, $N_\mathrm{c} = 1000, L = 30$.}\label{f:InstCache} %
\end{figure*}

This section demonstrates the instantaneous CCP of RR, LP, TLP, and LRU, and relate the results to the STF demonstrated in the preceding sections. 

The number of contents $N_\mathrm{c}$ and the cache size $L$ are set to 1000 and 30, respectively. For each of the four considered replacement schemes, the simulation consists of 5000 rounds. For each round, 2000 content requests are generated randomly based on a Zipf's distribution with parameter 0.8. The contents are sorted based on the request probability in a decreasing order. The cache is empty at the beginning. The instantaneous CCP for each content after each request is obtained and averaged over the 5000 rounds.

The resulting CCP for three selected contents, i.e., contents 1, 10, and 35, are shown in Fig.~\ref{f:InstCache}. It can be seen from Figs.~\ref{f:InstCacheRR}~and~\ref{f:InstCacheLRU} that, starting with an empty cache, RR and LRU becomes stationary faster than LP and TLP, which are shown in Fig.~\ref{f:InstCacheLES} and Fig.~\ref{f:InstCacheLEA}, respectively. In addition, by comparing Figs.~\ref{f:InstCacheRR}~and~\ref{f:InstCacheLRU}, it can be seen that LRU caches popular contents, e.g., content 1, with larger probabilities than RR. This is consistent with the observation from comparing Fig.~\ref{f:LRU3D} and Fig.~\ref{f:RR3D1}. The impact of $\alpha$ on the performance of LP can be seen from Fig.~\ref{f:InstCacheLES}, while the difference between TLP-A and TLP-P can be seen from Fig.~\ref{f:InstCacheLEA}. In the cases of LP and TLP, the cache hit probability of content 35 first increases and then decreases to zero. This corresponds to the `curvy' paths in the STF as shown in Fig.~\ref{f:LES3D} and Fig.~\ref{f:LEA3D}.

\section{Conclusion}

We have revisited the problem of modeling and analyzing cache replacement schemes under IRM with the objective of providing a rigorous yet intuitive general model from a novel perspective. Through this work, we have developed a basic tool set based on STF to characterize and illustrate cache replacement schemes. Our investigation has also been targeted at revealing insights regarding the relation between content popularity, knowledge of content popularity exploited by replacement schemes, and the resulting STFs. The model and methodology we have established in this paper can also be applied to multi-level cache and cache networks after appropriate extensions.  

\vspace{-2mm}
\appendix
\vspace{-2mm}

\subsection{Proof of Theorem~\ref{t:PropertyRRsteady}}\label{p:PropertyRRsteady}
\vspace{-1mm}
We first prove, using the STF, that the steady state is independent on $\phi$. It can be seen from eq.~\eqref{e:ul_RR} that $\phi$ is just a scaling factor in $u_{m, l, \mathrm{RR}}$. Moreover, the scaling factor is the same for any state $m$ and content $l$. Therefore, we can define a base STF such that at point $\boldsymbol{\eta}$ it satisfies:
\begin{align}\label{e:ul_RRBase}
\bar{u}_{m, l, \mathrm{RR}} (\boldsymbol{\eta}) =  \left\{
\begin{array}{ll}
\sum\limits_{\{k| m \in \mathcal{H}_{k,l}\}} \eta_k, \; &\text{if}\;  l \in \mathcal{C}_m \\ 
-L \eta_m, \;\; &\text{otherwise}. 
\end{array}
\right.
\end{align}
Then, it is easy to show that:
\begin{subequations}
	\begin{align} 
	\mathbf{u}_{l, \mathrm{RR}} (\boldsymbol{\eta}) &= \phi  \bar{\mathbf{u}}_{l, \mathrm{RR}} (\boldsymbol{\eta}), \\
	\mathbf{u}_{\mathrm{RR}} (\boldsymbol{\eta}) &= \phi  \bar{\mathbf{u}}_{\mathrm{RR}} (\boldsymbol{\eta}). 
	\end{align}
\end{subequations}
Accordingly, a change in $\phi$ can change the strength of the STF but does not alter the direction of the STF at any point in the state transition domain. Therefore, the steady state of RR must be independent on $\phi$.      

Next, we prove the property of the steady state. Based on the definition of STF in eq.~\eqref{e:StateTranFielddef}, the STF at the steady state SCP $\boldsymbol{\eta}^\star$ must be equal to $\mathbf{0}$. It follows that:
\begin{align}\label{e:uSetToZeroT1p}
u_{m, \mathrm{RR}} (\boldsymbol{\eta}^\star) =&  \sum\limits_{l \in \mathcal{C}_m} \upsilon_l \cdot u_{m, l, \mathrm{RR}} (\boldsymbol{\eta}^\star)  + \sum\limits_{l \notin \mathcal{C}_m} \upsilon_l \cdot u_{m, l, \mathrm{RR}} ( \boldsymbol{\eta}^\star ) \nonumber \\ 
= & \sum\limits_{l \in \mathcal{C}_m} \upsilon_l \phi \!\! \sum\limits_{\{k| m \in \mathcal{H}_{k,l}\}} \!\! \eta_k^\star + \sum\limits_{l \notin \mathcal{C}_m}  \upsilon_l (-L \phi)  \eta_m^\star      \nonumber \\
= & \,0,
\end{align}
which must hold for any $m \in \mathcal{S}$. Based on the definition of neighbors and content-specific neighbors, it can be seen that: 
\begin{align}\label{e:EqvSumT1p}
\sum\limits_{l \in \mathcal{C}_m} \upsilon_l \sum\limits_{\{k| m \in \mathcal{H}_{k,l}\}} \eta_k^\star = \sum\limits_{k \in \mathcal{H}_m }  \upsilon_{e(m, k)} \eta_{k}^\star.
\end{align}
Combining eq.~\eqref{e:EqvSumT1p} and eq.~\eqref{e:uSetToZeroT1p} gives eq.~\eqref{e:etaSteadyRNProp2}. \hfill $\blacksquare$

\subsection{Proof of Lemma~\ref{l:2ndEigenValueLESLEA}}\label{p:2ndEigenValueLESLEA}

First, we will prove that the second largest eigenvalue of both $\boldsymbol{\Theta}_\mathrm{LP}$ and $\boldsymbol{\Theta}_\mathrm{TLP}$ is the $(N_\mathrm{s} - 1, N_\mathrm{s} - 1)$th element. In the case of LP, the sum probability of state $k$ transitioning into any other state is given by $\alpha \sum_{l \in \mathcal{C}_{\bar{k}^\uparrow} }  \upsilon_l$, which is non-increasing with $k$. Accordingly, $\boldsymbol{\Theta}_\mathrm{LP}(m, m) \geq \boldsymbol{\Theta}_\mathrm{LP}(k, k)$ if $m > k$. Similarly, the same result can be shown for the TLP.   

Second, as the states are sorted based on the sum predicted request probability of their cached contents, it can be seen that $e(N_\mathrm{s}, N_\mathrm{s} - 1)$ is the $(N_\mathrm{c} -L + 1)$th content. Based on eq.~\eqref{e:ThetaOverallDefLES}, it can be seen that  $\boldsymbol{\Theta}_\mathrm{LP}(N_\mathrm{s} - 1, N_\mathrm{s} - 1)$ is equal to $1 - \alpha \tilde{\upsilon}_{\hat{l}}$ with $\hat{l}$ denoting $N_\mathrm{c} -L + 1$. Similarly,  $\boldsymbol{\Theta}_\mathrm{TLP}(N_\mathrm{s} - 1, N_\mathrm{s} - 1)$ is equal to $1 -  \tilde{\upsilon}_{\hat{l}}\phi_{\hat{l}, \hat{l} -1, N_\mathrm{s}-1} $ based on eq.~\eqref{e: ThetaOverallDefLEA}. \hfill$\blacksquare$

\subsection{Proof of Theorem~\ref{t:etaSteadyLRU}}\label{p:etaSteadyLRU}
The STF at the steady state SCP $\boldsymbol{\eta}^\star$ must be equal to $\mathbf{0}$. It follows that:
\begin{align}\label{e:uSetToZeroT2p}
u_{m, \mathrm{LRU}} (\boldsymbol{\eta}^\star) =&  \!\sum\limits_{l \in \mathcal{C}_m} \! \upsilon_l \cdot u_{m, l, \mathrm{LRU}} (\boldsymbol{\eta}^\star)  + \! \sum\limits_{l \notin \mathcal{C}_m} \upsilon_l \cdot u_{m, l, \mathrm{LRU}} ( \boldsymbol{\eta}^\star ) \nonumber \\ 
= & \sum\limits_{l \in \mathcal{C}_m} \upsilon_l   \sum\limits_{\{k| m \in \mathcal{H}_{k,l}\}}  \rho^\mathrm{LRU}_{e(k,m)|k} \eta_k^\star - \sum\limits_{l \notin \mathcal{C}_m}  \upsilon_l  \eta_m^\star        \nonumber \\
= & \, 0,
\end{align}
which must hold for any $m \in \mathcal{S}$. It can be shown that: 
\begin{align}\label{e:EqvSumT2p}
\sum\limits_{l \in \mathcal{C}_m} \upsilon_l \! \sum\limits_{\{k| m \in \mathcal{H}_{k,l}\}}  \rho^\mathrm{LRU}_{e(k,m)|k}  \eta_k^\star = \!\sum\limits_{k \in \mathcal{H}_m }  \upsilon_{e(m, k)}  \rho^\mathrm{LRU}_{e(k,m)|k} \eta_{k}^\star.
\end{align}
Combining eq.~\eqref{e:EqvSumT2p} and eq.~\eqref{e:uSetToZeroT2p} gives eq.~\eqref{e:etaSteadyLRU}. \hfill $\blacksquare$


{\footnotesize
\begin{thebibliography}{99}



\bibitem{J_ZPiao2019}
Z. Piao, M. Peng, Y. Liu, and M. Daneshmand, ``Recent Advances of Edge Cache in Radio Access Networks for Internet of Things: Techniques, Performances, and Challenges,'' \textit{IEEE Internet Things J.}, vol.~6, no.~1, pp.~1010--1028, Feb.~2019.

\bibitem{J_EMarkakis2017}
E. K. Markakis, K. Karras, A. Sideris, G. Alexiou, and E. Pallis, ``Computing, Caching, and Communication at the Edge: The Cornerstone for Building a Versatile 5G Ecosystem,'' \textit{IEEE Commun. Mag.}, vol.~55, no.~11, pp.~152--157, Nov.~2017.

\bibitem{J_MTang2018}
M. Tang, L. Gao, and J. Huang, ``Enabling Edge Cooperation in Tactile Internet via 3C Resource Sharing,''  \textit{IEEE J. Sel. Areas Commun.}, vol.~36, no.~11, pp.~2444--2454, Nov.~2018.


\bibitem{J_SZhang2018}
S. Zhang, P. He, K. Suto, P. Yang, L. Zhao, and X. Shen, ``Cooperative Edge Caching in User-Centric Clustered Mobile Networks,'' {\it IEEE Trans. Mobile Comput.}, vol.~17, no.~8, pp.~1791--1805, Aug.~2018.

\bibitem{J_MEmara2018}
M. Emara, H. Elsawy, S. Sorour, S. Al-Ghadhban, M. Alouini, and T.~Y.~Al-Naffouri, ``Optimal Caching in 5G Networks With Opportunistic Spectrum Access,'' \textit{IEEE Trans. Wireless Commun.}, vol.~17, no.~7, pp.~4447--4461, July~2018.

\bibitem{J_TXVu2018}
T. X. Vu, S. Chatzinotas, B. Ottersten, and T. Q. Duong, ``Energy Minimization for Cache-Assisted Content Delivery Networks With Wireless Backhaul,'' \textit{IEEE Wireless Commun. Lett.}, vol.~7, no.~3, pp.~332--335, June 2018.

\bibitem{J_GLee2014}
G. Lee, I. Jang, S. Pack, and X. Shen, ``FW-DAS: Fast Wireless Data Access Scheme in Mobile Networks,'' \textit{IEEE Trans. Wireless Commun.}, vol.~13, no.~8, pp.~4260--4272, Aug.~2014.


\bibitem{M_EBastug2014}
E. Bastug, M. Bennis, and M. Debbah, ``Living on the Edge: The Role of Proactive Caching in 5G Wireless Networks,'' \textit{IEEE Commun. Mag.}, vol.~52, no.~8, pp.~82--89, Aug.~2014.


\bibitem{J_KNDoan2016}
K. N. Doan, T. Van Nguyen, T. Q. S. Quek, and H. Shin, ``Content-Aware Proactive Caching for Backhaul Offloading in Cellular Network,'' \textit{IEEE Trans. Wireless Commun.}, vol.~17, no.~5, pp.~3128--3140, May 2018.

\bibitem{C_JGao2018}
J. Gao, L. Zhao, and L. Sun, ``Probabilistic Caching as Mixed Strategies in Spatially-Coupled Edge Caching,'' in \textit{Proc. 29th Biennial Symp.  Commun.}, Toronto, Canada, 2018.

\bibitem{J_JQiao2016}
J. Qiao, Y. He, and X. Shen, ``Proactive Caching for Mobile Video Streaming in Millimeter Wave 5G Networks,'' \textit{IEEE Trans. Wireless Commun.},  vol.~15, no.~10, pp.~7187--7198, Oct.~2016.


\bibitem{J_SOSomuyiwa2018}
S. O. Somuyiwa, A. Gy\"{o}rgy, and D. G\"{u}nd\"{u}z, ``A Reinforcement-Learning Approach to Proactive Caching in Wireless Networks,'' \textit{IEEE J. Sel. Areas Commun.}, vol.~36, no.~6, pp.~1331--1344, June~2018.


\bibitem{J_RPedarsani2016}
R. Pedarsani, M. A. Maddah-Ali, and U. Niesen, ``Online Coded Caching,'' \textit{IEEE/ACM Trans. Netw.}, vol.~24, no.~2, pp.~836--845, Apr.~2016.

\bibitem{J_SPodlipnig2003}
S. Podlipnig and L. B\"{o}sz\"{o}rmenyi,  ``A Survey of Web Cache Replacement Strategies,'' \textit{ACM Comput. Surv.}, vol.~35, no.~4, pp.~374--398, Dec.~2003.

\bibitem{J_LBelady1966}
L. A. Belady, ``A Study of Replacement Algorithms for a Virtual-Storage Computer,'' \textit{IBM Sys. J.}, vol.~5, no.~2, pp.~78--101, 1966.

\bibitem{J_DSleator1985}
D. D. Sleator and R. E. Tarjan, ``Amortized Efficiency of List Update and Paging Rules,''  \textit{Commun. ACM}, vol.~28 , no.~2, pp.~202--208, Feb.~1985.

\bibitem{J_GRao1978}
G.~S.~Rao, ``Performance Analysis of Cache Memories,''  \textit{J. ACM}, vol.~25, no.~3, pp.~378-395, July~1978.

\bibitem{J_AKarlin2000}
A. R. Karlin,, S. J. Phillips, and P. Raghavan, ``Markov Paging,'' \textit{SIAM J. Comput.}, vol.~30, no.~3, pp.~906-922, Aug.~2000. 

\bibitem{J_RHirade2010}
R.~Hirade and T.~Osogami. ``Analysis of Page Replacement Policies in the Fluid Limit,'' \textit{Operations Research}, vol.~58, no.~4, pp.~971-–984, July~2010.

\bibitem{J_HGomaa2013}
H. Gomaa, G. G. Messier, C. Williamson, and R. Davies, ``Estimating Instantaneous Cache Hit Ratio Using Markov Chain Analysis,'' \textit{IEEE/ACM Trans. Netw.}, vol.~21, no.~5, pp.~1472--1483, Oct.~2013.

\bibitem{C_STarnoi2015}
S. Tarnoi, V. Suppakitpaisarn, W. Kumwilaisak, and Y. Ji, ``Performance Analysis of Probabilistic Caching Scheme using Markov Chains,'' in \textit{Proc. IEEE LCN}, Clearwater Beach, USA, 2015, pp.~46--54.

\bibitem{J_JLi2018}
J. Li, S. Shakkottai, J. C. S. Lui, and V. Subramanian, ``Accurate Learning or Fast Mixing? Dynamic Adaptability of Caching Algorithms,'' \textit{IEEE J. Sel. Areas Commun.}, vol.~36, no.~6, pp.~1314--1330, June 2018.

\bibitem{J_JGaoTMC2018}
J.~Gao, S.~Zhang, L.~Zhao, and X.~Shen, ``The Design of Dynamic Probabilistic Caching with Time-Varying Content Popularity,'' submitted to {\it IEEE Trans. Mobile Comput.}, under review.

\bibitem{J_LChang2013}
L. Chang, J. Pan, and M. Xing, ``Effective Utilization of User Resources in PA-VoD Systems with Channel Heterogeneity,''  \textit{IEEE J. Sel. Areas Commun.}, vol.~31, no.~9, pp.~227--236, Sept.~2013.

\bibitem{J_MFiore2011}
M. Fiore, C. Casetti, and C. Chiasserini, ``Caching Strategies Based on Information Density Estimation in Wireless Ad Hoc Networks,'' \textit{IEEE Trans. Veh. Technol.},  vol.~60, no.~5, pp.~2194--2208, June 2011.

\bibitem{J_MMeddeb2019}
M.~Meddeb, A.~Dhraief, A.~Belghith, T.~Monteil, K.~Drira, and H.~Mathkour, ``Least Fresh First Cache Replacement Policy for NDN-based IoT networks,'' \textit{Pervasive Mob. Comput.}, vol~52, pp.~60--70, Jan.~2019. 

\bibitem{J_Meybodi2017}
Z.~H.~Meybodi, J.~Abouei, and A.~H.~F.~Raouf, ``Cache Replacement Schemes based on Adaptive Time Window for Video on Demand Services in Femtocell Networks,'' {\it IEEE Trans. on Mobile Comput.},  vol.~18, no.~7, pp.~1476--1487, July~2019. 

\bibitem{J_NKamiyama2016}
N. Kamiyama, Y. Nakano, and K. Shiomoto, ``Cache Replacement Based on Distance to Origin Servers,'' \textit{IEEE Trans. Netw. Service Manag.}, vol.~13, no.~4, pp.~848--859, Dec.~2016.

\bibitem{J_AChattopadhyay2018}
A. Chattopadhyay, B. B\l{}aszczyszyn, and H. P. Keeler, ``Gibbsian On-Line Distributed Content Caching Strategy for Cellular Networks,'' \textit{IEEE Trans. Wireless Commun.}, vol.~17, no.~2, pp.~969--981, Feb.~2018.

\bibitem{J_ELeonardi2018}
E. Leonardi and G. Neglia, ``Implicit Coordination of Caches in Small Cell Networks Under Unknown Popularity Profiles,'' \textit{IEEE J. Sel. Areas Commun.}, vol.~36, no.~6, pp.~1276--1285, June 2018.



\bibitem{J_JGaoPartII2018}
J.~Gao, L.~Zhao, and X.~Shen, ``The Study of Caching via State Transition Field - the Case of Time-Varying Popularity,'' \textit{IEEE Trans. Wireless Commun.}, accepted.

\bibitem{J_GSPaschos2018}
G. S. Paschos, G. Iosifidis, M. Tao, D. Towsley, and G. Caire, ``The Role of Caching in Future Communication Systems and Networks,''  \textit{IEEE J. Sel. Areas Commun.}, vol.~36, no.~6, pp.~1111--1125, June~2018.


\bibitem{B_DLevin2008}
D. Levin, Y. Peres, E. Wilmer, {\it Markov Chains and Mixing Times}. American Mathematical Society, Providence, RI, USA, 2008.

\bibitem{LN_SArora}
S.~Arora. {\it Random walks, Markov Chains, and How to Analyse Them}. Lecture Notes, Department of Computer Science, Princeton University, 2013. 

\bibitem{J_SGWalker2011}
S.~G. Walker, ``Bounds for the Second Largest Eigenvalue of a Transition Matrix,''  \textit{Linear and Multilinear Algebra}, vol.~59, no.~7, pp.~755-760, Apr.~2011. 





%
%

%
%
%
%





%
%
%
%
%
%
%
%
%
%
%
%
%
%
%
%
%
%
%
%
%
%

\end{thebibliography} }
\end{document}